\documentclass[prd, aps]{revtex4}
\usepackage{epsfig, amssymb, latexsym, amsfonts, amsmath, amsthm}

\newcommand{\KI}{ {\mathcal{K}} }
\newcommand{\OP}{ {\mathcal{O}} }

\newcommand{\tw}{ {\text{tw}} }

\newcommand{\PRa}{ \mathcal{P} }
\newcommand{\HB}{ \mathcal{H} }
\newcommand{\HS}{ \mathbf{H} }

\newcommand{\Y}{ \mathcal{Y} }

\newcommand{\X}{ \mathbf{X} }
\newcommand{\x}{ \mathbf{x} }
\newcommand{\D}{\mathbf{d}}
\newcommand{\pd}{ \partial }

\newcommand {\PP}{{\mathbb{P}}}
\newcommand {\PZ}{{\mathbb{P}\mathbb{Z}}}
\newcommand {\PZi}[1]{{\mathbb{P}^{ #1 }\mathbb{Z}}}

\newcommand {\kln}[1]{\left( #1 \right)}
\newcommand {\Kln}[1]{\bigl( #1 \bigr)}
\newcommand {\KLn}[1]{\Bigl( #1 \Bigr)}
\newcommand {\KLN}[1]{\biggl( #1 \biggr)}

\newcommand {\KLNo}[1]{\biggl( #1 \biggr.}

\newcommand {\oKLN}[1]{\biggl. #1 \biggr)}

\newcommand {\kls}[1]{\left\{ #1 \right\}}

\newcommand {\KLSS}[1]{\Biggl\{ #1 \Biggr\}}

\newcommand {\KLSSo}[1]{\Biggl\{ #1 \Biggr.}

\newcommand {\oKLSS}[1]{\Biggl. #1 \Biggr\}}

\newcommand {\kle}[1]{\left[ #1 \right]}
\newcommand {\Kle}[1]{\bigl[ #1 \bigr]}
\newcommand {\KLe}[1]{\Bigl[ #1 \Bigr]}
\newcommand {\KLE}[1]{\biggl[ #1 \biggr]}
\newcommand {\KLEE}[1]{\Biggl[ #1 \Biggr]}

\newcommand {\KLEEo}[1]{\Biggl[ #1 \Biggr.}

\newcommand {\oKLEE}[1]{\Biggl. #1 \Biggr]}

\newcommand {\matel}[3]{\left< #1 \left|\; #2\, \right| #3 \right>}
\newcommand {\Matel}[3]{\bigl< #1 \bigl|\; #2\, \bigr| #3 \bigr>}

\newcommand {\PIpe}[1]{\Bigl|_{#1} \Bigr.}

\newcommand {\Z} {{\mathbb{Z}}}

\newcommand {\Id} {{\mathbb{I}}}
\newcommand {\im}{{\text{i}}}
\newcommand {\e}{{\text{e}}}

\newcommand {\lra} {\leftrightarrow}

\newcommand {\la} {\leftarrow}

\newcommand {\ra} {\rightarrow}

\newcommand {\tx} {\tilde x}

\begin{document}

\title{Complete twist-decomposition for non-local QCD vector operators in $x$-space }

\author{ J\"org Eilers }
\email{ eilers@itp.uni-leipzig.de }
\affiliation{ Center for Theoretical Studies and Institute of
Theoretical Physics, Leipzig University, Augustusplatz~10,
D-04109~Leipzig, Germany}

\author{ Bodo Geyer }
\email{ geyer@itp.uni-leipzig.de }
\affiliation{ Center for Theoretical Studies and Institute of
Theoretical Physics, Leipzig University, Augustusplatz~10,
D-04109~Leipzig, Germany}

\author{ Markus Lazar }
\email{ lazar@itp.uni-leipzig.de } \affiliation{
Max-Planck-Institut for Mathematics in Science, Leipzig,
Inselstr.~22-26, D-04103~Leipzig, Germany}

\date{\today}

%!p####################################################################################################################################
\begin{abstract}

\vspace*{0.1cm}

\noindent A general procedure is introduced allowing for the
complete, infinite twist decomposition of non-local vector
operators in QCD off the light-cone. The method is applied to the
operators $\bar\psi(x) \gamma_\mu \psi(-x)$ and $\bar\psi(x)
\sigma_{\mu\nu} x^\nu \psi(-x)$ as well as to their matrix
elements thereby determining all power (resp. target mass)
corrections being relevant for the related distribution
amplitudes.

\vspace*{0.25cm}


\end{abstract}
%!p####################################################################################################################################

\maketitle

%%%%%%%%%%%%%%%%%%%%%%%%%%%%%%%%%%%%%%%%%%%%%%%%%%%%%%%%%%%%%%%%%%%%%%%%%%%%%%%%%%%%%%%%%%%%%%%%%%%%%%%%%%%%%%%%%%%%%%%%%%%%%%%%%%%%%%
\section{Introduction}
%%%%%%%%%%%%%%%%%%%%%%%%%%%%%%%%%%%%%%%%%%%%%%%%%%%%%%%%%%%%%%%%%%%%%%%%%%%%%%%%%%%%%%%%%%%%%%%%%%%%%%%%%%%%%%%%%%%%%%%%%%%%%%%%%%%%%%
The dependence of the non-perturbative distribution amplitudes on
the momentum transfer $Q^2$ is very important in phenomenological
considerations in QCD. In the past it became clear that the
non-local light-cone expansion together with the renormalization
group equation is a suitable tool to determine this dependence.
Experimental applications are, for example, the parton
distributions in deep inelastic scattering (DIS), the non-forward
double distribution amplitudes in deeply virtual Compton
scattering (DVCS) and the (vector) meson wave functions.
Furthermore, it became obvious that all the phenomenological
quantities appearing in the above experimental settings are
related to the Compton amplitude for non-forward scattering of a
virtual photon off a hadron given by
%!p____________________________________________________________________________________________________________________________________
\begin{equation}
\label{CA_nonf}
 T_{\mu\nu}(P_i,Q; S_i) =
 {\text{i}} \int {\mathrm{d}}^4 \! x \;\; {\text{e}}^{{\text{i}} qx} \; \bigl< P_2,S_2 \bigl|\;
 RT \left[J_\mu(x/2) J_\nu(-x/2) \, \mathcal{S} \right] \, \bigr| P_1,S_1 \bigr> \; ,
\end{equation}
%!p____________________________________________________________________________________________________________________________________
where $P_1 (P_2)$ and $S_1 (S_2)$ are the momenta and spins of the
incoming (outgoing) hadrons, $Q^2=-q^2,\,q= q_2-q_1= P_1-P_2$
denotes the momentum transfer and $\cal S$ is the (renormalized)
$S-$matrix.

To find an adequate representation of the Compton amplitude in
terms of non-local operators on the light-cone one applies the
non-local operator product expansion \cite{AZ78,ZAV,MRGHD} to the
renormalized time-ordered product occurring in the matrix element
(\ref{CA_nonf}). This leads immediately to compact expressions for
the coefficient functions and the corresponding bilocal light-cone
operators. If the quark propagator near the light-cone is
approximated by its most singular parts one finds the well-known
expression \footnote{As usual, (anti)symmetrization is taken with
the corresponding factorial, e.g., $T_{(\mu\nu)} =
\frac{1}{2}(T_{\mu\nu}+T_{\mu\nu}),\,T_{[\mu\nu]} =
\frac{1}{2}(T_{\mu\nu}-T_{\mu\nu})$. }
%!p____________________________________________________________________________________________________________________________________
\begin{eqnarray}
\label{str_wick}
 T\left[ J_\mu\left( \kappa \, x \right)J_\nu\left( -\kappa \, x \right) \right] \,
&\approx&
 \left[ \frac{1}{2\pi^2\left( x^2 - \text{i} \epsilon \right)^2}
 + \frac{m^2}{8\pi^2\left( x^2 - \text{i} \epsilon \right) } \right]
 \Bigl( g_{\mu\nu} \, O\left( \kappa \, x, - \kappa \, x \right)
 - 2 \, x_{( \mu } O_{ \nu )} \left( \kappa \, x, - \kappa \, x \right)
 \\
\nonumber &&
 \qquad\qquad\qquad\qquad\qquad\qquad\qquad\quad
  - {\text{i}}\,{\epsilon_{\mu\nu}}^{\alpha\beta} \; x_{\alpha}
O_{\beta}^5\left( \kappa \, x, - \kappa \, x\right) \Bigr)
\\
\nonumber
&&
 - \; \frac{ { \text{i} } \, m }{4\pi^2\left( x^2
 - {\text{i}} \epsilon \right) } \;\, \Bigl( g_{\mu\nu} \, N\left( \kappa \, x, -
\kappa \, x \right) + M_{\left[ \mu\nu \right]} \left( \kappa \,
x, - \kappa \, x \right) \Bigr) \, .
\end{eqnarray}
%!p____________________________________________________________________________________________________________________________________
Thereby, the `centered' non-local chiral-even (axial) vector
operators $O_\mu^{(5)}\left( \kappa \, x, - \kappa x \, \right)$
and the corresponding scalar operators $O^{(5)}\left( \kappa \, x,
- \kappa \, x \right) : = x^\mu O_\mu^{(5)}\left( \kappa \, x, -
\kappa x \, \right) $ are given by the following (anti)symmetrized
operators (here, and in the following, corresponding phase factors
are suppressed)
%!p____________________________________________________________________________________________________________________________________
\begin{eqnarray}
\label{vector_op}
%\nonumber
 O_\mu\left( \kappa \, x, - \kappa \, x \right)
&=&
 :\!\bar\psi\left( \kappa \, x \right)\,\gamma_\mu \,\psi\left( -\kappa \, x \right)\!: - :\!\bar\psi\left( -\kappa \, x \right)\,
\gamma_\mu \,\psi\left( \kappa \, x \right)\!: \;,
\\
%\nonumber
 O_\mu^5\left( \kappa \, x, - \kappa \, x \right)
&=&
 :\!\bar\psi\left( \kappa \, x \right)\, \gamma^5 \gamma_\mu \,\psi\left( -\kappa \, x \right)\!: + :\!\bar\psi\left( -\kappa \, x
\right)\, \gamma^5 \gamma_\mu \,\psi\left( \kappa \, x \right)\!: \;,
\\
 O\left( \kappa \, x, - \kappa \, x \right)
&=&
 :\!\bar\psi\left( \kappa \, x \right)\,  x \!\!\! /\, \,\psi\left( -\kappa \, x \right)\!:
 - :\!\bar\psi\left( -\kappa \, x \right)\,  x
\!\!\! /\, \,\psi\left( \kappa \, x \right)\!: \; ,
\\
 O^5\left( \kappa \, x, - \kappa \, x \right)
&=&
 :\!\bar\psi\left( \kappa \, x \right)\, \gamma^5\, x \!\!\! /\, \,\psi\left( -\kappa \, x \right)\!:
 + :\!\bar\psi\left( -\kappa \, x \right)\, \gamma^5\, x
\!\!\! /\, \,\psi\left( \kappa \, x \right)\!: \; ,
\end{eqnarray}
%!p____________________________________________________________________________________________________________________________________
whereas the `centered' chiral-odd skew tensor operator $M_{\left[
\mu\nu \right]} \left( \kappa \, x, - \kappa \, x \right)$ and the
scalar operator $N\left( \kappa \, x, - \kappa \, x \right)$ are
given by the following (anti)symmetrized operators
%!p____________________________________________________________________________________________________________________________________
\begin{eqnarray}
\label{skew}
 M_{\left[ \mu\nu \right]} \left( \kappa \, x, - \kappa \, x \right)
&=&
 :\!\bar\psi\left( \kappa \, x \right) \, \sigma_{\mu\nu} \, \psi\left( -\kappa \, x \right)\! : - :\!\bar\psi\left( -\kappa \, x
\right) \, \sigma_{\mu\nu} \, \psi\left( \kappa \, x \right)\! :\;,
\\
 N\left( \kappa \, x, - \kappa \, x \right)
&=&
 :\!\bar\psi\left( \kappa \, x \right) \; \psi\left( -\kappa \, x \right)\! : + \, :\!\bar\psi\left( -\kappa \, x \right) \, \psi\left(
\kappa \, x \right)\!: \; .
\end{eqnarray}
%!p____________________________________________________________________________________________________________________________________
%In addition we may construct an axial skew tensor operator by
%!p____________________________________________________________________________________________________________________________________
%\begin{equation}
%\label{noch_einer}
%M^5_{\left[ \mu\nu \right]} \left( \kappa \, x, - \kappa \, x \right)
%=
% :\!\bar\psi\left( \kappa \, x \right) \, \gamma^5 \, \sigma_{\mu\nu} \,
% \psi\left( -\kappa \, x \right)\! : - :\!\bar\psi\left( -\kappa \, x
%\right) \, \gamma^5 \, \sigma_{\mu\nu} \, \psi\left( \kappa \, x \right)\! : \; .
%\end{equation}
%!p____________________________________________________________________________________________________________________________________

For scalar operators, like $O^{(5)}\kln{\kappa\,x,-\kappa\,x}$ and
$N\kln{\kappa\,x,-\kappa\,x}$, as well as operators which can be
derived from them, e.g.~by applying derivatives w.r.t.~$x$, the
complete twist decomposition has been studied previously (see,
Ref.~\cite{GLR01}). Now, we extend these considerations to
vector operators, like $O_\mu^{(5)}\kln{\kappa\,x,-\kappa\,x}$ and
$M_{\mu}\kln{\kappa\,x,-\kappa\,x}$, the latter being obtained
from the skew tensor operator
$M_{[\mu\nu]}\kln{\kappa\,x,-\kappa\,x}$ by `external' contraction
with $x^\nu$,
%!p____________________________________________________________________________________________________________________________________
\begin{eqnarray}
\label{def_Mmu}
 M_{\mu}\kln{\kappa\,x,-\kappa\,x}
 &:=&
 x^\nu \; M_{[\mu\nu]}\kln{\kappa\,x,-\kappa\,x} \; .
\end{eqnarray}
%!p____________________________________________________________________________________________________________________________________
Obviously, the operators $O_\mu^{(5)} \left( \kappa \, x , -
\kappa \, x \right) $ contribute to the leading light-cone
singularity being independent of the quark-mass $m$ and also to
the lower light-cone singularities being proportional to $m^2$.
The operator $M_{[\mu\nu]}\left( \kappa \, x, - \kappa \, x
\right)$ only contributes to quark-mass terms and is therefore not
present if only leading contributions are considered. In addition,
its complete twist decomposition is much more involved and will be
studied separately.

All the non-perturbative distribution amplitudes which appear in
the above mentioned physical processes are Fourier transforms of
matrix elements whose unique input are the light-cone operators
(\ref{vector_op}) -- (\ref{def_Mmu}). Accordingly, all interesting
evolution equations of the distribution amplitudes result from the
renormalization group equation of these operators. This is the
reason why we consider the operators themselves and not their
phenomenologically relevant matrix elements.

With growing accuracy of experimental data for the various parton
distribution amplitudes more and more terms of the expansion of
the Compton amplitude (\ref{CA_nonf}) w.r.t. the variable
$M^2/Q^2$ will become accessible, where M is the nucleon (or,
eventually, the meson) mass. These terms are of quite different
origin. The first source are radiative corrections and a second
source are target mass corrections resulting from higher twist
contributions.

According to Gross and Treiman \cite{GT71} the various
contributions of definite (geometric) twist $\tau$ ($=$ canonical dimension
$d$ minus spin $j$) are obtained by a decomposition of the local
components $\OP_{\Gamma\alpha_1\ldots\alpha_n}$ of the bilocal
operators $\OP_\Gamma\kln{\kappa\,x,-\kappa\,x}$ into irreducible
tensor representations of the Lorentz group. As is well-known,
these representations are given by traceless tensors which, in
addition, are characterized by their symmetry type under index
permutations. In the case of the Lorentz group $SO(3,1)$ there exist four
generic symmetry types which we call I, II, III and IV, and which
are characterized by corresponding (normalized) Young operators
$\Y^{i}_{[m]},\,i= 1,...,4,\,$ with patterns $[m]$ of $m$ boxes,
$[m] = (m), (m-1,1), (m-2,1,1)$ and $(m-2,2)$, respectively (for a
more detailed group theoretical characterization see, e.g.,
Refs.~\cite{GLR99,GLR00}). In case of scalar operators only the
first pattern corresponding to totally symmetric tensors is
relevant because $\Gamma = \mathbb{I}$ and all tensor indices are
to be contracted by the symmetric tensor $x^{\alpha_1}\cdots
x^{\alpha_m}$. In case of vector operators only the first two
patterns are relevant with the `free' index $\Gamma \equiv \mu$
being symmetrized (pattern of type I) or antisymmetrized
(pattern of type II) with the remaining indices $\alpha_i,\,
i = 1,\ldots, m-1 $. In the case of antisymmetric and symmetric
tensor operators of rank 2 also patterns III and IV,
respectively, have to be taken into account.

Using this purely group theoretical procedure the (finite) twist
decomposition of non-local light-cone operators in configuration
space has been performed in Refs.~\cite{GLR99,GLR00}. However, if
one wants to calculate target mass corrections stemming from
higher twist operators one is forced to consider the (infinite)
twist decomposition off the light-cone. This infinite
decomposition includes all trace terms which, after Fourier
transformation to momentum space, lead to contributions suppressed
by powers of $M^2/Q^2$.

The target mass corrections resulting from leading and,
eventually, next-to-leading twist contributions have first been
discussed by Nachtmann \cite{Nachtmann73} in unpolarized deep
inelastic scattering. Later on, this method has been applied to
polarized deep inelastic scattering in
Refs.~\cite{BE76,BE76_b,W77,MU80,KU95}. Another method for the
determination of target mass effects was first given by Georgi and
Politzer \cite{GP} for unpolarized deep inelastic scattering and
then extended to polarized scattering and to general electro-weak
couplings by Refs.~\cite{PR98,BT99}.

However, all the above mentioned articles treated only the leading
twist contributions since, up to now, a complete off-cone
decomposition for non-local operators carrying free tensor indices
in $x$-space has been possible only when their $n$th moments are
totally symmetric allowing for the application of the group
theoretical results of Bargmann and Todorov \cite{BT77}.
Especially, this holds for all scalar operators, see
Ref.~\cite{GLR01} for the infinite twist decomposition of such
objects. But once free indices are involved the non-trivial Young
patterns occur. Here, we solve that problem when one free tensor
index appears.

The paper is organized as follows. Section 2 is devoted to the
deduction of the complete off-cone twist decomposition of generic
vector operators thereby determining all power corrections in
$x$--space contained in these operators. In section 3 we apply
these results to the operators $O_\mu\kln{\kappa\,x,-\kappa\,x}$
and $M_\mu\kln{\kappa\,x,-\kappa\,x}$. In section 4 the forward
and non-forward matrix elements of these operators are taken
thereby representing our results in a manner which can be used for
further phenomenological considerations.

%%%%%%%%%%%%%%%%%%%%%%%%%%%%%%%%%%%%%%%%%%%%%%%%%%%%%%%%%%%%%%%%%%%%%%%%%%%%%%%%%%%%%%%%%%%%%%%%%%%%%%%%%%%%%%%%%%%%%%%%%%%%%%%%%%%%%%
\section{Complete twist-decomposition for generic vector operators in $x$-space}
%%%%%%%%%%%%%%%%%%%%%%%%%%%%%%%%%%%%%%%%%%%%%%%%%%%%%%%%%%%%%%%%%%%%%%%%%%%%%%%%%%%%%%%%%%%%%%%%%%%%%%%%%%%%%%%%%%%%%%%%%%%%%%%%%%%%%%

In this Chapter we solve the problem of the twist decomposition
for a generic vector operator. Thereby, we make use of the
polynomial technique which already has been applied in order to
complete the twist-decomposition on the light-cone
\cite{GLR99,GLR00}. By this method the totally symmetric part of
any (local) operator is (sometimes only partly) truncated by the
coordinates $x^{\alpha_i}$ (see, Eqs.~(\ref{compx_momo_ex_v2}) and
(\ref{compx_momm_ex_v2}) below). In the following, we first list
the necessary ingredients, then we briefly illustrate the method
in case of scalar operators, and then we generalize to generic
vector operators.

To begin with let us first perform a formal Fourier transformation
of the non-local operators $O_\mu\kln{\kappa\,x,-\kappa\,x}$ and
$M_{[\mu\nu]}\kln{\kappa\,x,-\kappa\,x}$ followed by an expansion
into local operators (here, and in the following, without changing
the notation we suppress anti-symmetrization w.r.t. $\kappa$ being
part of the definition of the operators (\ref{vector_op}) and
(\ref{skew})),
%____________________________________________________________________________________________________________________________________
\begin{eqnarray}
\label{compx_momo}
 O_\mu\kln{\kappa \, x,-\kappa\,x}
&=&
 \int \text{d}^4\!u \; \, O_\mu\kln{u} \; \e^{\im\kappa\,\kln{ux}}
 = \sum_{n=0}^\infty \, \frac{\kln{\im\kappa}^n}{n!} \; O_{\mu|n}\kln{x} \;,
\\
\label{compx_momm}
 M_{[\mu\nu]}\kln{\kappa \, x,-\kappa\,x}
&=&
 \int \text{d}^4\!u \; \, M_{[\mu\nu]}\kln{u} \; \e^{\im\kappa\,\kln{ux}}
 = \sum_{n=0}^\infty \, \frac{\kln{\im\kappa}^n}{n!} \; M_{[\mu\nu]|n}\kln{x} \; .
\end{eqnarray}
%!p____________________________________________________________________________________________________________________________________
(For notational simplicity we understand the factor $1/(2\pi)^4$
to be included into the measure of the Fourier transformation.)
%(obviously, this transformation has nothing to do with the Fourier
%transformation performed in the Compton amplitude (\ref{CA_nonf})).
Obviously, the operators (\ref{compx_momo}) and (\ref{compx_momm})
are homogeneous in $\kappa$, $u$ and $x$ whereas those which are
obtained after `external' multiplication with $x^\mu$ are not.
Therefore, the explicit expressions for the moments of the vector
operators (\ref{vector_op}) and (\ref{def_Mmu}) read
%p____________________________________________________________________________________________________________________________________
\begin{eqnarray}
\label{compx_momo_ex}
 O_{\mu|n}\kln{x}
&=&
 \int \text{d}^4\!u \; \, O_\mu\kln{u} \; \kln{ux}^n = \int \text{d}^4\!u \; \kln{\bar\psi \, \gamma_\mu \psi}\kln{u} \; \kln{ux}^n \; ,
\\
\label{compx_momm_ex}
 M_{\mu|n+1}\kln{x}
&=&
 \int \text{d}^4\!u \; \, M_{[\mu\nu]}\kln{u} \; x^\nu \kln{ux}^n = \frac{1}{n+1} \int \text{d}^4\!u \; \kln{\bar\psi \, \sigma_{\mu\nu}
\psi}\kln{u} \; \pd^\nu_u \kln{ux}^{n+1} \; .
\end{eqnarray}
%!p____________________________________________________________________________________________________________________________________
If we perform a local expansion of the operators without a
preceding Fourier transformation, the moments $O_{\mu|n}\kln{x}$
and $M_{\mu|n+1}\kln{x}$ receive the following (equivalent)
representation
%!p____________________________________________________________________________________________________________________________________
\begin{eqnarray}
\label{compx_momo_ex_v2}
 O_{\mu|n}\kln{x}
&=&
 \kln{-\im}^n \; x^{\alpha_1} \dots x^{\alpha_n} \; O_{\mu\alpha_1\dots\alpha_n} \; ,
\\
\label{compx_momm_ex_v2}
 M_{\mu|n+1}\kln{x}
&=&
 \kln{-\im}^n \; x^\nu x^{\alpha_1} \dots x^{\alpha_n} \; M_{[\mu\nu]\alpha_1\dots\alpha_n} \; ,
\end{eqnarray}
%!p____________________________________________________________________________________________________________________________________
with
%!p____________________________________________________________________________________________________________________________________
\begin{eqnarray}
\label{compx_momo_ex_v3}
 O_{\mu\alpha_1\dots\alpha_n}
 &=&
 \bar \psi\kln{0} \gamma_\mu \,
 \overset{\lra}{D}_{\alpha_1} \dots \overset{\lra}{D}_{\alpha_n} \psi\kln{0} \; ,
\\
\label{compx_momm_ex_v3}
 M_{\mu\nu\alpha_1\dots\alpha_n}
 &=&
 \bar \psi\kln{0} \sigma_{\mu\nu} \,
 \overset{\lra}{D}_{\alpha_1} \dots \overset{\lra}{D}_{\alpha_n} \psi\kln{0} \; .
\end{eqnarray}
%!p____________________________________________________________________________________________________________________________________
$\overset{\lra}{D}_{\alpha}$ is the covariant derivative,
%!p____________________________________________________________________________________________________________________________________
\begin{equation}
\label{covariant} \overset{\lra}{D}_{\alpha} =
\overset{\ra}{\partial_\alpha} - \overset{\la}{\partial_\alpha} +
2\im \, g \, A_\alpha\kln{x}\,,
\end{equation}
%!p____________________________________________________________________________________________________________________________________
with $g$ being the strong coupling constant and $A_\alpha\kln{x}$
the gluon field in the fundamental representation.
%The operators (\ref{compx_momo_ex_v2}) and (\ref{compx_momm_ex_v2})
%are only needed for supplementary reasons.

Let us begin with the construction of the infinite twist
decomposition for generic {\em scalar operators} $\OP\kln{\kappa
\, x,-\kappa\,x}$ with local components $\OP_n\kln{x}=
\kln{-\im}^n x^{\alpha_1}\ldots x^{\alpha_n}\, \OP_{\alpha_1\ldots
\alpha_n}$. Using the group theoretic results of Bargmann and
Todorov \cite{BT77} this decomposition reads
%!p____________________________________________________________________________________________________________________________________
\begin{equation}
\label{compx_bt1}
 \OP_n\kln{x} = \sum_{j=0}^{\kle{\frac{n}{2}}} \; c\kln{j,n}\; \kln{x^2}^j
 \; H_{n-2j}\kln{x^2,\Box} \; \Box^j \; \OP_n\kln{x}\,,
\end{equation}
%!p____________________________________________________________________________________________________________________________________
with the projection operator $ H_{n}\kln{x^2,\Box}$ onto traceless
homogeneous polynomials of degree $n$,
%!p____________________________________________________________________________________________________________________________________
\begin{eqnarray}
\label{compx_no3}
 H_{n}\kln{x^2,\Box}
 &=&
 \sum_{k=0}^{\kle{\frac{n}{2}}} \; d\kln{k,n} \; \kln{x^2}^k \Box^k
 \equiv
 \sum_{k=0}^{\kle{\frac{n}{2}}} \;
 \frac{ \kln{-1}^k \; \kln{n-k}! }{ 4^k \; k! \; n! } \; \kln{x^2}^k \Box^k \;
 ;
\end{eqnarray}
thereby, the corresponding coefficients are given by
\begin{eqnarray}
\label{compx_no1}
 c\kln{j,n}
 &=&
 \frac{ \kln{n+1-2j}! }{4^j \; j! \; \kln{n+1-j}! } \; ,
\\
\label{compx_no2}
 d\kln{k,n}
&=&
 \frac{ \kln{-1}^k \; \kln{n-k}! }{ 4^k \; k! \; n! } \; .
\end{eqnarray}
%!p____________________________________________________________________________________________________________________________________
The (dimensionless) projector $H_{n}\kln{x^2,\Box}$ is obtained as
the unique solution of the requirement of tracelessness of a
(scalar) homogeneous polynomial of degree $n$ according to
\begin{equation}
 \label{scalartrace}
 \Box\;H_{n}\kln{x^2,\Box} \;  \OP_n\kln{x} = 0\,.
\end{equation}

Formula (\ref{compx_bt1}) provides us with a separation of the
traceless operators of definite twist, $H_{n-2j}\kln{x^2,\Box} \,
\Box^j \; \OP_n\kln{x}$, from the trace terms, $\kln{x^2}^j$,
which are obtained by contracting the indices of the various products of
the metrics by $x^\alpha$'s. Since scalar operators always obey
symmetry type I, all traceless scalar operators are automatically
of definite twist and, therefore, no further symmetrization is
necessary. This key observation can be generalized also to
operators with free tensor indices (which however require
subsequent application of Young symmetrizers, see below).

With the notations (\ref{compx_no3}) -- (\ref{compx_no2}), after
exchanging the summations and changing summation indices from
$\kln{j,k}$ to $\kln{j,r=k+j}$, the result (\ref{compx_bt1}) can
be rewritten in the form
%!p____________________________________________________________________________________________________________________________________
\begin{equation}
\label{compx_bt2}
 \OP_n\kln{x} = %\underset{= \Id }{\underbrace{
 \sum_{r=0}^{\kle{\frac{n}{2}}} \; \kln{x^2}^r \Box^r \; \sum_{j=0}^{r} \; c\kln{j,n} \;
d\kln{r-j,n-2j} %}}
\; \,\OP_n\kln{x} \; .
\end{equation}
%!p____________________________________________________________________________________________________________________________________
Since the undecomposed operator $\OP_n\kln{x}$ occurs on both
sides of the equation the differential operator on the right hand
side must be equal to the identity $\Id$. This is only possible if
the coefficients $c$ and $d$ fulfill the equations
%!p____________________________________________________________________________________________________________________________________
\begin{equation}
\label{compx_inv}
 \sum_{j=0}^{r} \; c\kln{j,n} \; d\kln{r-j,n-2j}
 = \delta_{0r} \qquad \text{for} \qquad
 r=0,\dots,\kle{\frac{n}{2}}\;,
\end{equation}
%!p____________________________________________________________________________________________________________________________________
which can be checked by explicit calculation. In this sense, $c$
and $d$ are inverse to each other. However, if $c$ is regarded as
an unknown coefficient, equation (\ref{compx_inv}) can be used to
determine $c$ since $d$ is already defined by the
projection operator (\ref{compx_no3}).

Now, we generalize these observations to the case of the {\em
vector operators} under consideration, $O_\mu$ and $M_\mu$, which
generically will be denoted by $\OP_\mu$. First, in order to get
an ansatz which provides us with an analogous separation of the
trace terms from the traceless operators we have to specify which
kind of trace terms could appear. Obviously, all possible traces
of the operators $O_{\mu\alpha_1\dots\alpha_n}$ and
$M_{[\mu\nu]\alpha_1\dots\alpha_n}$ are multiplied by products of
metrics $g_{\beta_i \beta_j}$ with $\kls{\beta_i}=
\kls{\mu,\nu,\alpha_1,\dots,\alpha_n}$. After contraction with the
symmetric tensor $x^{\alpha_1} \dots x^{\alpha_n}$ these metrics
obtain the form of products of $x^2$, $x_{\mu}$, $x_{\nu}$ and
$g_{\mu\nu}$ depending on the number of $x^{\alpha}$'s being
contracted with the different $g_{\beta_i \beta_j}$. Since we are
dealing with vector operators or operators being antisymmetric in
$\mu$ and $\nu$ only two types of trace terms appear, namely,
$x^2$-traces and $x_\alpha$-traces.
%From now on we will denote the free index by $\alpha$.

Traces of type $x^2$ must be multiplied by a traceless vector
operator (carrying the free index $\alpha$) and traces of type
$x_{\alpha}$ must be multiplied by a traceless scalar operator.
Therefore, a preliminary ansatz reads
%!p____________________________________________________________________________________________________________________________________
\begin{equation}
\label{ansatz}
 \OP_{\alpha|n}\kln{x} = \sum_{j=0}^{\kle{\frac{n+1}{2}}}  \KLE{ \kln{x^2}^j \;
\HS_{\alpha|n-2j}^\rho\kln{x,\pd} + x_\alpha \,\kln{x^2}^{j-1}\;
\HS_{n+1-2j}^\rho\kln{x,\pd} } \; \OP_{\rho|n}\kln{x} \; ,
\end{equation}
%!p____________________________________________________________________________________________________________________________________
where $\HS_{\alpha|n-2j}^\rho\kln{x,\pd}$ is a projection operator
on traceless vector operators of order $n-2j$ and
$\HS_{n+1-2j}^\rho\kln{x,\pd}$ is a projection operator on scalar
traceless operators of order $n+1-2j$. The most general ansatz for
these projectors is
%!p____________________________________________________________________________________________________________________________________
\begin{eqnarray}
\label{compx_ans1}
 \HS_{\alpha|n-2j}^\rho\kln{x,\pd}
&=&
 \;\;\,\, c_1\kln{j,n} \cdot H_{\alpha|n-2j}^\rho\kln{x,\pd} \, \Box^j
\\
\nonumber
&&
 + \, c_2\kln{j,n} \cdot \partial_\alpha \, H_{n+1-2j}\kln{x^2,\Box} \; \Box^{j-1} \, \pd^\rho
\\
\nonumber
&&
 + \, c_3\kln{j,n} \cdot \partial_\alpha \, H_{n+1-2j}\kln{x^2,\Box} \; \Box^{j} \; x^\rho \; ,
\\
\nonumber
&&
\\
\label{compx_ans2}
 \HS_{n+1-2j}^\rho\kln{x,\pd}
&=&
 \;\;\,\, c_4\kln{j,n} \cdot H_{n+1-2j}\kln{x^2,\Box} \; \Box^{j-1} \, \pd^\rho
\\
\nonumber
&&
 + \, c_5\kln{j,n} \cdot H_{n+1-2j}\kln{x^2,\Box} \; \Box^{j} \; x^\rho \; ,
\end{eqnarray}
%!p____________________________________________________________________________________________________________________________________
where $H_{\alpha|n}^\rho\kln{x,\pd}$ is the (primary) projector
onto traceless homogeneous polynomials of degree $n$ bearing a
free Lorentz index and obeying the defining equations
\begin{eqnarray}
\label{vectortrace}
 \Box \, H_{\alpha|n}^\rho \; \OP_{\rho|n}\kln{x}
=
 0 \; ,
\qquad % \label{vectortrace}
 \partial^\alpha \, H_{\alpha|n}^\rho \; \OP_{\rho|n}\kln{x}
=
 0 \; ,
\end{eqnarray}
which has been determined already in Refs.~\cite{GLR99,GLR00},
%!p____________________________________________________________________________________________________________________________________
\begin{eqnarray}
 H^{\rho}_{\alpha|n}\kln{x,\pd}
&=& \kle{ \delta_{\alpha}^{\rho} - \frac{1}{\kln{n+1}^2} \KLn{ n
\, x_{\alpha} - \frac{1}{2} \; x^2 \, \partial_\alpha } \,
\partial^\rho } \, H_n\kln{x^2,\Box}
\\
\nonumber &=&  \delta_{\alpha}^{\rho} \, H_n\kln{x^2,\Box} -
\frac{1}{n\,\kln{n+1}^2} \; \x_\alpha\,H_{n-1}\kln{x^2,\Box}\,
\D^\rho \, .
\end{eqnarray}
%!p____________________________________________________________________________________________________________________________________
Here, the set of traceless vector operators is enlarged by
differentiated scalar traceless operators, which is allowed since
a differentiation does not destroy the condition of tracelessness.
Namely, by construction, $\HS_{\alpha|n-2j}^\rho\;
\OP_{\rho|n}\kln{x}$ and $\HS_{n+1-2j}^\rho\; \OP_{\rho|n}\kln{x}$
fulfill the equations (in the following, the arguments of the
projectors $\HS$ and $H$ will be omitted)
%!p____________________________________________________________________________________________________________________________________
\begin{eqnarray}
 \Box \, \HS_{\alpha|n-2j}^\rho \; \OP_{\rho|n}\kln{x}
=
 0 \; ,
\qquad
 \partial^\alpha \, \HS_{\alpha|n-2j}^\rho \; \OP_{\rho|n}\kln{x}
=
 0 \; ,
\qquad
 \Box \, \HS_{n+1-2j}^\rho \; \OP_{\rho|n}\kln{x}
=
 0 \;.
\end{eqnarray}
%!p____________________________________________________________________________________________________________________________________

Above, for later convenience we introduced the operators
\begin{eqnarray}
 \x_\alpha &=& x_{\alpha} (x\partial +1) \,
 - \hbox{\large$\frac{1}{2}$} \; x^2 \,
 \partial_\alpha \,,\\
 \D_\alpha &=& (x\partial +1)\partial_\alpha
 - \hbox{\large$\frac{1}{2}$} x_\alpha \, \Box\,,
\end{eqnarray}
with the properties
\begin{eqnarray}
\label{xx}
 \x_\alpha\,H_n\kln{x^2,\Box} &=&
 H_{n+1}\kln{x^2,\Box}x_\alpha\,(x\partial +1),
 \\
 \label{dd}  (x\partial +1)\,\partial_\alpha\, H_n\kln{x^2,\Box} &=&
 H_{n-1}\kln{x^2,\Box}\D_\alpha\,.
\end{eqnarray}
%when acting on a homogeneous polynomial of order $n$.
%_________________________________________________________________________________________________________________________________
%
Using these relations it is easily seen that also the following
equation holds,
\begin{equation}
\label{compx_rela}
 x^\beta \, H^{\rho}_{\beta|n} = H_{n+1} \; x^\rho \; ,
\end{equation}
%!p____________________________________________________________________________________________________________________________________
reflecting the fact that a traceless vector operator remains
traceless once it is contracted with $x$. Let us also introduce
the generators of angular-momentum, $\X_{[\mu\nu]} = - x_{[\mu} \,
\pd_{\nu]}$, and of dilations, $\X = x\partial +1$, which commute
with $H_n$,
\begin{equation}
\label{H_commute}
 \Kle{\X , H_n } = 0, \qquad
 \Kle{\X_{[\mu\nu]} , H_n } = 0\,.
\end{equation}
$\D_\sigma$ is a generalization of the interior derivative on the
light-cone which has been introduced by Bargman and Todorov
\cite{BT77}. It is nilpotent on the light-cone but off the
light-cone the square is proportional to $x^2$:
%!p____________________________________________________________________________________________________________________________________
\begin{equation}
\D^{\sigma} \D_\sigma = \hbox{\large$\frac{1}{4}$} \; x^2 \;
\Box^2 \; .
\end{equation}
%!p
Let us remark, that the operators $\X,\, x_\mu,\, \D_\mu$ and
$\X_{[\mu\nu]}$ obey the conformal algebra also off the
light-cone.

The set of unknown objects is now reduced to the coefficients
$c_1$ to $c_5$ which can be determined by the same argument as has
been used in the scalar case to fix the coefficient $c$. We
therefore expect five equations similar to (\ref{compx_inv}),
which have to be solved iteratively in $r$. As a preparatory step
we insert the explicit expressions for $H_{\alpha|n}^\rho$ and
$H_n$ into the equations (\ref{compx_ans1}) and (\ref{compx_ans2})
and perform the necessary differentiations followed by a change of
summation indices from $\kln{j,k}$ to $\kln{j,r}$. The result for
the five coefficients $c_1(j,n)$ to $c_5(j,n)$ is:
%\begin{itemize}
%\item
%{ Operator related to $c_1$:
%!p____________________________________________________________________________________________________________________________________
\begin{eqnarray}
\label{compx_eqn1}
&&
 \sum_{j=0}^{\kle{\frac{n}{2}}} \kln{x^2}^j \; c_1\kln{j,n} \, H_{\alpha|n-2j}^\rho \; \Box^j \; \OP_{\rho|n}\kln{x}
\\
\nonumber && \quad\qquad
 = \sum_{r=0}^{\kle{\frac{n}{2}}} \;
\sum_{j=0}^r \; c_1\kln{j,n} \cdot d\kln{r-j,n-2j} \; \frac{
\kln{x^2}^{r-1} }{ \kln{n+1-2j}^2 }
\\
\nonumber && \qquad \quad\qquad
 \times \KLEEo{ \KLn{ r-j + \kln{n+1-2j}^2 } \; x^2 \, \delta_{\alpha}^\rho \, \Box } - 2 \kln{r-j} \kln{n+1-r-j} \; x_\alpha \,
x^\rho \, \Box
\\
\nonumber
&&
 \qquad \qquad \quad \qquad
 - \kln{ n-r-j } \; x^2 \; x_\alpha \, \partial^\rho \; \Box + \kln{r-j} \; x^2 \; x^\rho \, \partial_\alpha \; \Box \oKLEE{
\; + \; \frac{1}{2} \, \kln{x^2}^2 \; \partial_\alpha \,
\partial^\rho \; \Box } \Box^{r-1} \; \OP_{\rho|n}\kln{x} \; ,
%\end{eqnarray}
%!p____________________________________________________________________________________________________________________________________
%}
%\item
%{ Operator related to $c_2$:
%!p____________________________________________________________________________________________________________________________________
%\begin{eqnarray}
\\
\label{compx_eqn2}
&&
 \sum_{j=0}^{\kle{\frac{n+1}{2}}} \kln{x^2}^j \, c_2\kln{j,n}  \; \partial_\alpha \; H_{n+1-2j} \; \partial^\rho \, \Box^{j-1}
  \; \OP_{\rho|n}\kln{x} \\
\nonumber
 && \quad\qquad
= \sum_{r=0}^{\kle{\frac{n+1}{2}}} \; \sum_{j=0}^{r} c_2\kln{j,n} \cdot d\kln{r-j,n+1-2j} \; \kln{x^2}^{r-1}
\KLE{ 2 \kln{r-j} \cdot x_\alpha \,
\partial^\rho + x^2 \; \partial_\alpha \, \partial^\rho } \Box^{r-1} \; \OP_{\rho|n}\kln{x} \; ,
%\end{eqnarray}
%!p____________________________________________________________________________________________________________________________________
%}
%\item
%{ Operator related to $c_3$:
%!p____________________________________________________________________________________________________________________________________
%\begin{eqnarray}
\\
\label{compx_eqn3}
&&
 \sum_{j=0}^{\kle{\frac{n+1}{2}}} \kln{x^2}^j \, c_3\kln{j,n}  \; \partial_\alpha \; H_{n+1-2j} \; \Box^{j} \; x^\rho
 \; \OP_{\rho|n}\kln{x}\\
\nonumber && \quad\qquad
 = \sum_{r=0}^{\kle{\frac{n+1}{2}}} \;
\sum_{j=0}^{r} c_3\kln{j,n} \cdot d\kln{r-j,n+1-2j} \;
\kln{x^2}^{r-1}
\\
\nonumber && \qquad \quad\qquad
 \times \KLEEo{} x^2 \, \delta_{\alpha}^\rho \, \Box + 2 \kln{r-j} \cdot x_\alpha \, x^\rho \, \Box
+ 4 \, r \kln{r-j} \cdot x_\alpha \, \partial^\rho + x^2 \, x^\rho \, \partial_\alpha \, \Box + 2 \, r \cdot x^2 \;
\partial_\alpha \, \partial^\rho \oKLEE{} \Box^{r-1} \; \OP_{\rho|n}\kln{x} \; ,
%\end{eqnarray}
%!p____________________________________________________________________________________________________________________________________
%}
%\item
%{ Operator related to $c_4$:
%!p____________________________________________________________________________________________________________________________________
%\begin{eqnarray}
\\\label{compx_eqn4}
&&
 \sum_{j=0}^{\kle{\frac{n+1}{2}}} \kln{x^2}^{j-1} \, c_4\kln{j,n} \; x_\alpha \, H_{n+1-2j} \; \partial^\rho \; \Box^{j-1}
  \; \OP_{\rho|n}\kln{x}\\
 \nonumber
 && \quad\qquad
= \sum_{r=0}^{\kle{\frac{n+1}{2}}} \; \sum_{j=0}^{r} c_4\kln{j,n} \cdot d\kln{r-j,n+1-2j} \; \kln{x^2}^{r-1}
\KLe{ x_\alpha \, \partial^\rho } \Box^{r-1} \; \OP_{\rho|n}\kln{x} \; ,
%\end{eqnarray}
%!p____________________________________________________________________________________________________________________________________
%}
%\item
%{ Operator related to $c_5$:
%!p____________________________________________________________________________________________________________________________________
%\begin{eqnarray}
\\
\label{compx_eqn5}
&&
 \sum_{j=0}^{\kle{\frac{n+1}{2}}} \kln{x^2}^{j-1} \, c_5\kln{j,n} \; x_\alpha \, H_{n+1-2j} \; \Box^{j} \; x^\rho
  \; \OP_{\rho|n}\kln{x}\\
 \nonumber
 && \quad\qquad
 = \sum_{r=0}^{\kle{\frac{n+1}{2}}} \;
\sum_{j=0}^{r} c_5\kln{j,n} \cdot d\kln{r-j,n+1-2j} \; \kln{x^2}^{r-1}
\KLe{ x_\alpha \, x^\rho \; \Box \; + \; 2\,r \cdot x_\alpha \, \partial^\rho }
\;\Box^{r-1} \; \OP_{\rho|n}\kln{x} \; .
\end{eqnarray}
%!p____________________________________________________________________________________________________________________________________
%}
%\end{itemize}
Looking onto these equations we observe five independent tensor
structures of dimension zero, % (\ref{compx_eqn3})
%!p____________________________________________________________________________________________________________________________________
\begin{eqnarray}
\label{compx_str}
 (x^2)^{r-1} \left\{ \;x^2 \, \delta_{\alpha}^\rho \, \Box \, ,\quad
 x_\alpha \, x^\rho \, \Box \,  ,\quad
 x_\alpha \, \partial^\rho \,  ,\quad
 x^2 \, x^\rho \, \partial_\alpha \, \Box \,  ,\quad
 x^2 \; \partial_\alpha \, \partial^\rho \;\right\} \Box^{r-1}\,,
\end{eqnarray}
%!p____________________________________________________________________________________________________________________________________
which should be taken as basis of reference operators.
%Let us
%remark that an operator of type (\ref{compx_eqn3}) exists for any
%operator with $m$ free indices and always contains the maximum
%number of possible tensor structures which can be build out of the
%indices
%$\kls{\sigma_k}=\kls{\alpha_1,\dots,\alpha_m,\rho_1,\dots,\rho_m}$
%and the objects $x_{\sigma_k}$, $\pd_{\sigma_k}$ and
%$\delta^{\sigma_k}_{\sigma_l}$. The second property is also true
%for the operator (\ref{compx_eqn1}) but this type of object may
%not exist for operators with more then two free indices.
%
Now we collect all terms belonging to the structures
(\ref{compx_str}) by imposing the constraint
%!p____________________________________________________________________________________________________________________________________
\begin{equation}
 \sum_{j=0}^{\kle{\frac{n+1}{2}}}  \kln{ \kln{x^2}^j \; \HS_{\alpha|n-2j}^\rho +
 x_\alpha \,\kln{x^2}^{j-1}\; \HS_{n+1-2j}^\rho } \overset{!}{=}
\delta_{\alpha}^\rho\,,
\end{equation}
%!p____________________________________________________________________________________________________________________________________
which again reflects the fact that the trace decomposition
(\ref{ansatz}) must be a formal representation of
$\delta_{\alpha}^\rho$. As expected we find a set of five
equations which, in the order of the reference operators, read:
%!p><><><><><><><><><><><><><><><><><><><><><><><><><><><><><><><><><><><><><><><><><>
%\begin{itemize} \item
%{ Equation related to the operator $x^2 \, \delta_{\alpha}^\rho \, \Box $:
%!p____________________________________________________________________________________________________________________________________
\begin{align}
\label{compx_eqnc1}
 & \sum_{j=0}^r
 \KLSS{ c_1\kln{j,n}  d\kln{r-j,n-2j}  \frac{\kln{n+1-2j}^2 + r - j}{ \kln{n+1-2j}^2 } \;
 + \; c_3\kln{j,n} d\kln{r-j,n+1-2j} } = \delta_{0r}\,,
%\end{eqnarray}
%!p____________________________________________________________________________________________________________________________________
%}
%!p><><><><><><><><><><><><><><><><><><><><><><><><><><><><><><><><><><><><><><><><><><
%\item
%{ Equation related to the operator $x_\alpha x^\rho \, \Box$:
%!p____________________________________________________________________________________________________________________________________
\\ %\begin{eqnarray}
\label{compx_eqnc2}
 &
 \sum_{j=0}^r \KLSSo{} c_1\kln{j,n}  d\kln{r-j,n-2j}
  \frac{ 2\kln{r-j}\kln{n+1-r-j} }{\kln{n+1-2j}^2}
%\\ \nonumber& \qquad \qquad
- \; d\kln{r-j,n+1-2j} \KLE{2\kln{r-j} c_3\kln{j,n}
  + c_5\kln{j,n} } \oKLSS{} = 0\,,
%\end{eqnarray}
%!p____________________________________________________________________________________________________________________________________
%}
%!p><><><><><><><><><><><><><><><><><><><><><><><><><><><><><><><><><><><><><><><><><><><><><><><><><><><><><><><><><><><><><><><><><><
%\item
%{ Equation related to the operator $ x^2 \; x_\alpha \partial^\rho \; \Box $:
%!p____________________________________________________________________________________________________________________________________
\\ %\begin{eqnarray}
\label{compx_eqnc3}
&
 \sum_{j=0}^{r-1} c_1\kln{j,n}  d\kln{r-1-j,n-2j}  \frac{ n-r-1-j }{\kln{n+1-2j}^2}
\\
\nonumber
& \qquad
 - \sum_{j=0}^{r} d\kln{r-j,n+1-2j} \KLE{ 2\kln{r-j}  c_2\kln{j,n}
 + 4\,r\kln{r-j}  c_3\kln{j,n} + c_4\kln{j,n} + 2\,r c_5\kln{j,n} } =
 0\,,
%\end{eqnarray}
%!p____________________________________________________________________________________________________________________________________
%}
%!p><><><><><><><><><><><><><><><><><><><><><><><><><><><><><><><><><><><><><><><><><><><><><><><><><><><><><><><><><><><><><><><><><><
%\item
%{ Equation related to the operator $ x^2 \; x^\rho \partial_\alpha \; \Box $:
%!p____________________________________________________________________________________________________________________________________
\\ %\begin{eqnarray}
\label{compx_eqnc4}
 & \sum_{j=0}^r \KLSS{ c_1\kln{j,n}  d\kln{r-j,n-2j}  \frac{ r-j }{\kln{n+1-2j}^2} \;
 + \; c_3\kln{j,n} d\kln{r-j,n+1-2j} } = 0\,,
%\end{eqnarray}
%!p____________________________________________________________________________________________________________________________________
%}
%!p><><><><><><><><><><><><><><><><><><><><><><><><><><><><><><><><><><><><><><><><><><><><><><><><><><><><><><><><><><><><><><><><><><
%\item
%{ Equation related to the operator $ \kln{x^2}^2 \; \partial_\alpha \partial^\rho \; \Box $:
%!p____________________________________________________________________________________________________________________________________
\\ %\begin{eqnarray}
\label{compx_eqnc5} &
 \sum_{j=0}^{r-1} c_1\kln{j,n}  d\kln{r-1-j,n-2j}  \frac{1}{2\kln{n+1-2j}^2}
%\\ \nonumber&& \qquad
 + \sum_{j=0}^{r} d\kln{r-j,n+1-2j} \KLE{ c_2\kln{j,n} + 2\,r  c_3\kln{j,n} } =
 0\,.
\end{align}
%!p____________________________________________________________________________________________________________________________________
%}
%!p><><><><><><><><><><><><><><><><><><><><><><><><><><><><><><><><><><><><><><><><><><><><><><><><><><><><><><><><><><><><><><><><><><
%\end{itemize}
%!p><><><><><><><><><><><><><><><><><><><><><><><><><><><><><><><><><><><><><><><><><><><><><><><><><><><><><><><><><><><><><><><><><><
The shift $r \ra r-1$ in the first term of equations
(\ref{compx_eqnc3}) and (\ref{compx_eqnc5}) is due to the fact
that the third and the fifth of the structures (\ref{compx_str})
are shifted by $r \ra r+1$ in equation (\ref{compx_eqn1}). For
$r=0$ these two sums are empty. This shift must be compensated
otherwise $c_1$ would not contribute to the correct power of $r$.

The set of equations (\ref{compx_eqnc1}) -- (\ref{compx_eqnc5})
can be regarded as a generalization of equation (\ref{compx_inv}).
It can be solved iteratively starting with $r=j=0$. We find
$c_1\kln{0,n}=1$ and  $c_i\kln{0,n}=0,\, i=2, ...,5$. Then, we
insert this solution into the set of linear equations for $r=1$.
The resulting system is easily solved and the solution determines
$c_1\kln{2,n}$ to $c_5\kln{2,n}$, and so on.
This iterative procedure leads to the following general solution
for $c_1$ to $c_5$:
%!p____________________________________________________________________________________________________________________________________
\begin{alignat}{3}
\label{compx_solc1}
 c_1\kln{j,n}
&=&
 \frac{ \kln{n+1-2j}! }{4^j \, j! \, \kln{n+1-j}! }
 &=& c\kln{j,n} \; ,
\\
\label{compx_solc2}
 c_2\kln{j,n}
&=&
 -\, \frac{ 2j \; \kln{n+1-2j}! }{4^j \, j! \, \kln{n+3-2j} \, \kln{n+1-j}! }
 &=& \,-\, \frac{2j}{ n+3-2j } \cdot c\kln{j,n} \; ,
\\
\label{compx_solc3}
 c_3\kln{j,n}
&=&
 \frac{ j \, \kln{n+1-2j}! }{4^j \, j! \,\kln{n+3-2j} \, \kln{n+2-j}! }
 &=& \frac{1}{ 2\kln{n+2-j}}\, \frac{2j}{ n+3-2j } \cdot c\kln{j,n} \; ,
\\
\label{compx_solc4}
 c_4\kln{j,n}
&=&
 \frac{ 4j \, \kln{n+2-2j}! }{4^{j} \, j! \,\kln{n+3-2j} \, \kln{n+1-j}! }
 &=&\, 2\, \kln{ n+2-2j } \, \frac{2j}{ n+3-2j } \cdot c\kln{j,n} \; ,
\\
\label{compx_solc5}
 c_5\kln{j,n}
&=&
 \,- \,\frac{ 2j \; \kln{n+2-2j}! }{4^j \, j! \, \kln{n+3-2j} \, \kln{n+2-j}! }
 &=&\, - \,\frac{ \kln{ n+2-2j } }{ \kln{n+2-j} }\, \frac{2j}{ n+3-2j }
\cdot c\kln{j,n} \; .
\end{alignat}
%!p____________________________________________________________________________________________________________________________________

For a direct proof of the above solution we should have a closer
look at (\ref{compx_eqnc1}) and (\ref{compx_eqnc4}). A subtraction
of both equations yields a relation for $c_1\kln{j,n}$
%!p____________________________________________________________________________________________________________________________________
%\begin{equation}
% \sum_{j=0}^{r} \; c_1\kln{j,n} \; d\kln{r-j,n-2j} = \delta_{0r}
% \qquad \text{for} \qquad r=0,\dots,\kle{\frac{n}{2}} \; ,
%\end{equation}
%!p____________________________________________________________________________________________________________________________________
which coincides with Eq.~(\ref{compx_inv}). $c_1\kln{j,n}$ is
therefore identical to $c\kln{j,n}$ given by
Eq.~(\ref{compx_no1}). The proof of the remaining solutions
(\ref{compx_solc2}) to (\ref{compx_solc5}) for $c_2\kln{j,n}$ to
$c_5\kln{j,n}$ is obtained by induction. Let us show this for
$c_3$. Namely, inserting $c_1$ into Eq.~(\ref{compx_eqnc4}) it is
easily seen that this equation is valid for $r=0$ and for $r=1$.
Now, assume that (\ref{compx_solc3}) is valid for $c_3(r,n)$, then
Eq.~(\ref{compx_eqnc4}) is chosen for $r+1$ and resolved w.r.t.
$c_3(r+1,n)$,
%!p____________________________________________________________________________________________________________________________________
\begin{align}
 c_3\kln{r+1,n} = - \sum_{j=0}^{r} \, \frac{c\kln{j,n} \, d\kln{r+1-j,n-2j} }{n+1-2j}
 \kln{ \frac{r+1-j}{n+1-2j} +
 \frac{j\,\kln{n-r-j}}{\kln{n+2-j}\kln{n+3-2j}} } \; .
 \nonumber
\end{align}
%!p____________________________________________________________________________________________________________________________________
Performing this sum we find the desired solution for
$c_3\kln{r+1,n}$,
%!p____________________________________________________________________________________________________________________________________
\begin{align}
 c_3\kln{r+1,n} = \frac{\kln{r+1}\,\Kln{n+1-2\kln{r+1}}!}{4^{r+1}\,
 \Kln{n+3-2\kln{r+1}}\,\kln{r+1}!\,\Kln{n+2-\kln{r+1}}! } \; .
 \nonumber
\end{align}
%!p____________________________________________________________________________________________________________________________________
One can now continue to prove by induction also the remaining
three results for $c_2$, $c_4$ and $c_5$. We omit the explicit
calculations since they yield no additional information.

Now, inserting the solution for $c_1$ to $c_5$ into the projectors
(\ref{compx_ans1}) and (\ref{compx_ans2}) of the decomposition
(\ref{ansatz}) we get the complete {\em trace decomposition} as
follows
%!p____________________________________________________________________________________________________________________________________
\begin{eqnarray}
\label{compx_trace}
 \OP_{\alpha|n}\kln{x}
&=&
 \sum_{j=0}^{ \kle{ \frac{n+1}{2} } } \frac{ \kln{n+1-2j}! }{ 4^j \, j! \, \kln{n+1-j}! } \, \kln{x^2}^{j-1} \KLEEo{} x^2 \;
H_{\alpha|n-2j}^\rho \; \Box
 \\ \nonumber &&
\qquad\qquad\qquad\qquad\qquad\qquad\qquad
  + \frac{4j}{(n+3-2j)(n+2-j)}\, \x_\alpha\,  H_{n+1-2j} \,
 \D^\rho  \oKLEE{} \Box^{j-1} \OP_{\rho|n}\!\kln{x} \,
.
\end{eqnarray}
%!p____________________________________________________________________________________________________________________________________
This is the generalization of the scalar trace decomposition
(\ref{compx_bt1}) to vector operators carrying a free index
$\alpha$.

In the scalar case the trace decomposition already is equal to the
infinite twist decomposition because traceless scalar operators
always have well-defined twist. However, once free indices are
involved we need to apply suitable projection operators onto the
desired symmetry type. As has been mentioned before there exist
two different symmetry types I and II when only one free index
occurs.
The corresponding Young projection operators have to be adjusted
to the polynomial method. They have been determined in
Refs.~\cite{GLR99,GLR01} and are given by
%!p____________________________________________________________________________________________________________________________________
\begin{eqnarray}
\label{compx_young1}
 \Y^{1\;\alpha}_{\mu|n}
&=&
 \frac{1}{n+1} \; \pd_\mu \; x^\alpha \; ,
\\
\label{compx_young2}
 \Y^{2\;\alpha}_{\mu|n}
&=&
 \frac{2}{n+1} \; x^\sigma \; \delta_{[\mu}^\alpha \, \pd_{\sigma]} \; .
\end{eqnarray}
%!p____________________________________________________________________________________________________________________________________
As is easily proven, $\Y^{1\;\alpha}_{\mu|n}$ and
$\Y^{2\;\alpha}_{\mu|n}$ fulfill the correct projection
properties,
%!p____________________________________________________________________________________________________________________________________
\begin{eqnarray}
\label{compx_proprop1}
 \sum_{k=1}^2 \Y^{k\;\alpha}_{\mu|n} \;\, \OP_{\alpha|n}\kln{x}
&=&
 \delta_{\mu}^{\alpha} \;\, \OP_{\alpha|n}\kln{x}  ,
\\
\label{compx_proprop2}
 \Y^{k\;\mu}_{\nu|n} \; \Y^{l\;\alpha}_{\mu|n} \;\, \OP_{\alpha|n}\kln{x}
&=&
 \delta_l^k \; \Y^{l\;\alpha}_{\nu|n} \;\, \OP_{\alpha|n}\kln{x} \; .
\end{eqnarray}
%!p____________________________________________________________________________________________________________________________________

 Now, we must have a closer look at the trace decomposition
(\ref{compx_trace}) in order to see where the representation
(\ref{compx_proprop1}) of $\delta_{\mu}^{\alpha}$ must be
inserted:\\
(a) In the second term of the expansion, behind the terms
$(x^2)^{j-1}\x_\alpha$, only traceless scalar operators occur
belonging to symmetry type I. Therefore, they have already
well-defined twist which is not destroyed by applying
$\partial_\alpha$.
%Despite this term can be regarded as fully
%twist decomposed we indicate the `suppressed' symmetry
%projector by the identity $\Id$.
\\
(b) In the first term, however, the operator $H_{\alpha|n-2j}^\rho
\, \Box^j \, \OP_{\rho|n}\kln{x}$, despite being traceless, has no
well-defined twist because it has no definite symmetry behavior
and, therefore, here we must insert the decomposition
(\ref{compx_proprop1}).\\
(c) Furthermore, the expansion (\ref{compx_trace}) must have a
definite overall-symmetry. So we have to multiply the whole
expression with the decomposition of $\delta_\mu^\alpha$.

Following these arguments we find
%!p____________________________________________________________________________________________________________________________________
\begin{eqnarray}
\label{compx_trace_sym}
 \OP_{\mu|n}\kln{x}
&=&
 \sum_{k=1}^2 \Y^{k\;\alpha}_{\mu|n} \;
 \sum_{j=0}^{ \kle{ \frac{n}{2} } } \frac{ \kln{n+1-2j}! }{ 4^j \, j! \, \kln{n+1-j}! } \,
\kln{x^2}^{j-1} \KLEEo{} x^2 \; \sum_{l=1}^2
\Y^{l\;\beta}_{\alpha|n-2j} \; H_{\beta|n-2j}^\rho \; \Box
\\
\nonumber &&\qquad\qquad\qquad\qquad\qquad\qquad\qquad\qquad
 + \frac{4j}{(n+3-2j)(n+2-j)}\, \x_\alpha  \; %\Id \;
 H_{n+1-2j}\,
 \D^\rho  \oKLEE{}\Box^{j-1} \OP_{\rho|n}\!\kln{x} .
\end{eqnarray}
%!p____________________________________________________________________________________________________________________________________
Let us remark that the symmetry projectors
$\Y^{k\;\alpha}_{\mu|n}$ and $\Y^{l\;\beta}_{\alpha|n}$ in
(\ref{compx_trace_sym}), although being of the same form
(\ref{compx_young1}) and (\ref{compx_young2}), respectively,
depending on their symmetry type $k$ and $l$ have a different
interpretation. If a symmetry projector acts on a traceless
operator it generates an operator of well-defined twist whereas
the global projectors $\Y^{k\;\alpha}_{\mu|n}$ fix the symmetry of
the undecomposed operator. According to this, the symmetry
properties of the operators of well-defined twist are nested
within the decomposition of the different global Young patterns
being determined by appropriate symmetry projectors.

To find the infinite twist-decomposition in $x$-space we must now
apply all symmetry projectors $\Y^{k\;\alpha}_{\mu|n}$. This is
done by calculating the following expressions,
%!p____________________________________________________________________________________________________________________________________
\begin{equation}
\label{compx_term1}
\begin{array}{lll}
\Y^{k\;\alpha}_{\mu|n} \; \kln{x^2}^{j-1} \x_{\alpha} \; H_{n+1-2j} & \text{ for } & k=1,2\,,
 \\ \\
\Y^{k\;\alpha}_{\mu|n} \; \kln{x^2}^{j}
\Y^{l\;\beta}_{\alpha|n-2j} \; H^{\rho}_{\beta|n-2j} & \text{ for
} & k,l=1,2\,,
\end{array}
\end{equation}
%!p____________________________________________________________________________________________________________________________________
in terms of operators of well-defined twist. We begin with $k=1$
which corresponds to symmetry type I. The calculation is almost
trivial and we find
%!p____________________________________________________________________________________________________________________________________
\begin{eqnarray}
%\label{compx_term_k1_1}
% \Y^{1\;\alpha}_{\mu|n} \; \kln{x^2}^{j} \partial_{\alpha} \; H_{n+1-2j}
%&=&
% \frac{n+1-2j}{n+1} \; \pd_\mu \, \kln{x^2}^j \; H_{n+1-2j} \; ,
%\\
\label{compx_term_k1_2}
 \Y^{1\;\alpha}_{\mu|n} \; \kln{x^2}^{j-1} \x_{\alpha} \; H_{n+1-2j}
&=&
 \frac{n+3-2j}{2\,(n+1)} \; \pd_\mu \, \kln{x^2}^j \; H_{n+1-2j} \; ,
\\
\label{compx_term_k1_3}
 \Y^{1\;\alpha}_{\mu|n} \; \kln{x^2}^{j} \Y^{1\;\beta}_{\alpha|n-2j} \; H^{\rho}_{\beta|n-2j}
&=&
 \frac{1}{n+1} \; \pd_\mu \, \kln{x^2}^j \; H_{n+1-2j} \; x^\rho \; ,
\\
\label{compx_term_k1_4}
 \Y^{1\;\alpha}_{\mu|n} \; \kln{x^2}^{j} \Y^{2\;\beta}_{\alpha|n-2j} \; H^{\rho}_{\beta|n-2j}
&=&
 0 \; ,
\end{eqnarray}
%!p____________________________________________________________________________________________________________________________________
where we have used the relation (\ref{compx_rela}). Inserting
these results into the expansion (\ref{compx_trace_sym}) we find
the infinite twist-decomposition for vector operators of symmetry
type I as it already has been given in Ref.~\cite{GLR01}, Chapter
III.C,
%!p____________________________________________________________________________________________________________________________________
\begin{equation}
\label{compx_twist_sym1}
 \Y^{1\;\rho}_{\mu|n} \; \OP_{\rho|n}\kln{x} = \frac{1}{n+1} \, \sum_{j=0}^{ \kle{ \frac{n+1}{2} } } \frac{ \kln{n+2-2j}! }{ 4^j \, j! \,
\kln{n+2-j}! } \; \pd_\mu \, \kln{x^2}^{j} \; H_{n+1-2j} \; \Box^j \; x^\rho \; \OP_{\rho|n}\kln{x} \; .
\end{equation}
%!p____________________________________________________________________________________________________________________________________
This result also shows, that the trace decomposition
(\ref{compx_trace}) fulfills the consistency condition $x^\alpha
\, \OP_{\alpha|n}\kln{x} = \OP_{n+1}\kln{x}$ which must hold since
an infinite twist-decomposition of symmetry type I always reduces
to the infinite twist decomposition of a scalar operator for any
number of free indices. All indices are then freed from their
contraction with the coordinates by partial differentiations, see
Ref.~\cite{GLR01}, page 121, for the details.

The calculation for $k=2$ is a little bit more involved but leads
straightforwardly to the following result:
%!p____________________________________________________________________________________________________________________________________
\begin{eqnarray}
%\label{compx_term_k2_1}
% \Y^{2\;\alpha}_{\mu|n} \; \kln{x^2}^{j} \partial_{\alpha} \; H_{n+1-2j}
%&=&
% \frac{4\,j}{n+1} \; \kln{x^2}^{j-1} x^\sigma \, \X_{[\mu\sigma]} \; H_{n+1-2j} \; ,
%\\
\label{compx_term_k2_2}
 \Y^{2\;\alpha}_{\mu|n} \; \kln{x^2}^{j-1} \x_{\alpha} \; H_{n+1-2j}
&=&
 - \frac{2\,(n+2-j)}{n+1} \; \kln{x^2}^{j-1} x^\sigma \, \X_{[\mu\sigma]} \; H_{n+1-2j} \; ,
\\
\label{compx_term_k2_3}
 \Y^{2\;\alpha}_{\mu|n} \; \kln{x^2}^{j} \Y^{1\;\beta}_{\alpha|n-2j} \; H^{\rho}_{\beta|n-2j}
&=&
 \frac{4\,j}{\kln{n+1}\kln{n+1-2j}} \; \kln{x^2}^{j-1} x^\sigma \, \X_{[\mu\sigma]} \; H_{n+1-2j} \; x^\rho \; ,
\\
\label{compx_term_k2_4}
 \Y^{2\;\alpha}_{\mu|n} \; \kln{x^2}^{j} \Y^{2\;\beta}_{\alpha|n-2j} \; H^{\rho}_{\beta|n-2j}
&=&
 \frac{2}{n+1-2j} \; \kln{x^2}^j \; x^\sigma \kln{ \delta_{[\mu}^\rho \, \pd_{\sigma]} + \frac{1}{n+1-2j} \; \X_{[\mu\sigma]} \;
\pd^\rho } H_{n-2j} \; .
\end{eqnarray}
%!p____________________________________________________________________________________________________________________________________
Again, we have used the relation (\ref{compx_rela}).
Inserting these results %(\ref{compx_term_k2_2}) -- (\ref{compx_term_k2_4})
into the expansion (\ref{compx_trace_sym})
we find the following decomposition:
%!p____________________________________________________________________________________________________________________________________
\begin{eqnarray}
\label{compx_twist_symII}
 \Y^{2\;\rho}_{\mu|n} \; \OP_{\rho|n}\kln{x}
&=&
 \sum_{j=0}^{ \kle{ \frac{n+1}{2} } }
 \frac{ \kln{n+1-2j}! }{ 4^j \, j! \, \kln{n+1-j}! } \,
 \Bigg[
 \kln{x^2}^{j} x^\sigma
 \frac{2}{n+1-2j}  \kln{ \delta_{[\mu}^\rho \, \pd_{\sigma]} +
 \frac{1}{n+1-2j} \; \X_{[\mu\sigma]} \; \pd^\rho }\, H_{n-2j} \,\Box^j\, \OP_{\rho|n}
\nonumber\\
 &&
 \qquad
 + \, \frac{ 8j \kln{n+2-2j}}{(n+3-2j)(n+1-2j)}
 \kln{x^2}^{j-1} x^\sigma \,\X_{[\mu\sigma]}\, H_{n+1-2j} \,
 \kln{ \frac{1}{n+1} \;
 \Box^j x^\rho - \Box^{j-1}\,\partial^\rho }\OP_{\rho|n}
 \Bigg]\,.
\end{eqnarray}

At this point we introduce the {\em complete set of local
operators of well-defined twist}
which appear in the resulting infinite twist-decomposition:\\
First we define {\em two scalar operators} which appear behind
$x_\mu$-traces,
%!p____________________________________________________________________________________________________________________________________
%\begin{center}
%\fbox{
%\begin{minipage}[c]{10cm}
\begin{align}
\label{compx_twist_op1}
 \OP^{\tw\kln{\tau_0+2j}\;\text{A}}_{n+1-2j}\kln{x}
&:=
 H_{n+1-2j} \; \Box^j \; x^\rho \; \OP_{\rho|n}\kln{x} \; ,
\\
\label{compx_twist_op2}
 \OP^{\tw\kln{\tau_0+2j}\;\text{B}}_{n+1-2j}\kln{x}
&:=
 H_{n+1-2j} \; \Box^{j-1} \; \pd^\rho \; \OP_{\rho|n}\kln{x} \;,
 \qquad \quad j\ge 1.
\\ %\end{eqnarray}
%\end{minipage} }
%\end{center}
%!p____________________________________________________________________________________________________________________________________
\intertext{Secondly, we define {\em two vector operators} which
are deduced from the previous two scalar operators
%$\OP^{\tw\kln{\tau_0+2j}\;\text{A}}_{n+1-2j}\kln{x}$ and
%$\OP^{\tw\kln{\tau_0+2j}\;\text{B}}_{n+1-2j}\kln{x}$,
by a differentiation $\pd_\mu$ and multiplication with the
pre-factor $1/\kln{n+1-2j}$, which is included for later
convenience,}
%\\ \nonumber \intertext{ }
%!p____________________________________________________________________________________________________________________________________
%\begin{center}
%\fbox{
%\begin{minipage}[c]{12.5cm}
%\begin{eqnarray}
\label{compx_twist_op1_vector_def}
 \OP^{\tw\kln{\tau_0+2j}\;\text{A}}_{\mu|n-2j}\kln{x}
&:=
 \frac{1}{n+1-2j} \; \pd_\mu \; \OP^{\tw\kln{\tau_0+2j}\;\text{A}}_{n+1-2j}\kln{x}
\\
\label{compx_twist_op1_vector} & =
 \frac{1}{\kln{n+1-2j}^2} \; H_{n-2j} \; \D_\mu \; \Box^j \; x^\rho \; \OP_{\rho|n}\kln{x} \; ,
\\
\label{compx_twist_op2_vector_def}
 \OP^{\tw\kln{\tau_0+2j}\;\text{B}}_{\mu|n-2j}\kln{x}
&:=
 \frac{1}{n+1-2j} \; \pd_\mu \; \OP^{\tw\kln{\tau_0+2j}\text{B}}_{n+1-2j}\kln{x}
\\
\label{compx_twist_op2_vector} & =
 \frac{1}{\kln{n+1-2j}^2} \; H_{n-2j} \; \D_\mu \; \Box^{j-1} \; \pd^\rho \;
 \OP_{\rho|n}\kln{x} \;,
\qquad\qquad j\ge 1\;.
\\ %\end{eqnarray}
%\end{minipage} }
%\end{center}
%!p____________________________________________________________________________________________________________________________________
\intertext{Finally, we define an {\em antisymmetric tensor
operator} and a {\em related vector operator},}
%!p____________________________________________________________________________________________________________________________________
%\begin{center}
%\fbox{
%\begin{minipage}[c]{17cm}
%\begin{eqnarray}
\label{compx_twist_op3}
 \OP^{\tw\kln{\tau_0+1+2j}\;\text{C}}_{[\mu\sigma]|n-1-2j}\kln{x}
&:=
 \frac{2}{n+1-2j} \kln{ \delta_{[\mu}^\rho \, \pd_{\sigma]} + \frac{1}{n+1-2j} \; \X_{[\mu\sigma]} \; \pd^{\rho} } \; H_{n-2j} \;
\Box^{j} \; \OP_{\rho|n}\kln{x} %\nonumber
\\
\label{compx_twist_op3b} &=
 \frac{2}{\kln{n-2j}\kln{n+1-2j}} \; H_{n-1-2j} \kln{ \delta_{[\mu}^\rho \, \D_{\sigma]} + \frac{1}{n+1-2j} \; \X_{[\mu\sigma]} \;
\D^{\rho} } \; \Box^{j} \; \OP_{\rho|n}\kln{x} \; ,
\\
\label{compx_twist_op3_vec}
 \OP^{\tw\kln{\tau_0+1+2j}\;\text{C}}_{\mu|n-2j}\kln{x}
&:=
 x^\sigma \; \OP^{\tw\kln{\tau_0+1+2j}\;\text{C}}_{[\mu\sigma]|n-1-2j}\kln{x} \; .
\end{align}
%\end{minipage} }
%\end{center}
%!p____________________________________________________________________________________________________________________________________
Obviously, these scalar, vector and tensor operators
%(\ref{compx_twist_op1_vector_def}) -- (\ref{compx_twist_op3_vec})
vanish for $j >\kle{ \frac{n+1}{2} }, \kle{ \frac{n}{2} }$  and
$\kle{ \frac{n-1}{2} }$, respectively, by construction.

Here, $\tau_0$ denotes the {\em minimal twist} of the operator
$\OP_{\mu}\kln{\kappa\,x, - \kappa\,x}$, given by its canonical
dimension $d$ minus its spin $j$, thereby also observing external
contractions with $x^\mu$ and/or $\partial^\mu$:\\
%!p____________________________________________________________________________________________________________________________________
%\begin{equation}
% \tau_0 = \text{canonical dimension} \kln{d} - \text{number of free indices} \kln{i} - \text{number of external $x^\mu$ contractions}
%\kln{c} \; .
%\end{equation}
%!p____________________________________________________________________________________________________________________________________
(i) For the vector operator $O_\mu\kln{\kappa\,x, - \kappa\,x}$ we
find $\tau_0=2$. Then, the first four operators
(\ref{compx_twist_op1}) -- (\ref{compx_twist_op2_vector_def}) are
of even twist beginning with twist-2 and the last two operators of
type C are of odd twist beginning with twist-3 .\\
(ii) For the operator $M_\mu\kln{\kappa\,x, - \kappa\,x}$ we find
$\tau_0 =1$. In that case operators of type
(\ref{compx_twist_op1}) and (\ref{compx_twist_op1_vector_def}) do
not exist due to the `internal' antisymmetry of $M_\mu$ not
allowing symmetry type I
which is related to operators of type A. \\
It is important to remark that $\tau_0$ only enumerates the twist
content of various operators but has no major impact on the twist
decomposition itself.
%
%Depending on its `external' structure, an operator of twist
%$\tau_0$ must not necessarily exist, for example the twist of
%$\tau_0=2$ is realized for $O_\mu$ but a twist of $\tau_0=1$ does
%expansion (\ref{compx_twist_sym2}) for $j=0$) and is therefore zero.

The first two scalar operators,
$\OP^{\tw\kln{\tau_0+2j}\;\text{A}}_{n+1-2j}\kln{x}$ and
$\OP^{\tw\kln{\tau_0+2j}\;\text{B}}_{n+1-2j}\kln{x}$, lie in the
tensor space $\mathbf{T}\kln{\frac{n+1-2j}{2},\frac{n+1-2j}{2}}$
while the two related vector operators,
$\OP^{\tw\kln{\tau_0+2j}\;\text{A}}_{\mu|n-2j}\kln{x}$ and
$\OP^{\tw\kln{\tau_0+2j}\;\text{B}}_{\mu|n-2j}\kln{x}$, lie in the
space $\mathbf{T}\kln{\frac{n-2j}{2},\frac{n-2j}{2}}$. The forth
operator,
$\OP^{\tw\kln{\tau_0+1+2j}\;\text{C}}_{[\mu\sigma]|n-1-2j}\kln{x}$,
lies in the space $\mathbf{T}\kln{\frac{n-2j}{2},\frac{n-2-2j}{2}}
\oplus \mathbf{T}\kln{\frac{n-2-2j}{2},\frac{n-2j}{2}}$ and the
related vector operator in
$\mathbf{T}\kln{\frac{n+1-2j}{2},\frac{n-1-2j}{2}} \oplus
\mathbf{T}\kln{\frac{n-1-2j}{2},\frac{n+1-2j}{2}}$. The four
operators of type A and B correspond to symmetry type I while the
tensor or vector operators (\ref{compx_twist_op3}) and
(\ref{compx_twist_op3_vec}) correspond to symmetry type II.

The definition of the operators (\ref{compx_twist_op1}) --
(\ref{compx_twist_op3_vec}) is motivated by the polynomial
technique. Namely, as already has been explained the terms
proportional to $x^2$ and $x_\alpha$ are trace terms being
multiplied by traceless operators. It is therefore convincing to
include the derivative $\pd_\mu$ together with $1/\kln{n+1-2j}$ in
(\ref{compx_twist_op1_vector_def}) and
(\ref{compx_twist_op2_vector_def}) into the definition of the
vector operators
$\OP^{\tw\kln{\tau_0+2j}\;\text{A}}_{\mu|n-2j}\kln{x}$ and
$\OP^{\tw\kln{\tau_0+2j}\;\text{B}}_{\mu|n-2j}\kln{x}$. With the
same argument we include $x^\sigma$ into the definition of
$\OP^{\tw\kln{\tau_0+1+2j}\;\text{C}}_{\mu|n-2j}\kln{x}$. The
vector operators obey the relations
%!p____________________________________________________________________________________________________________________________________
\begin{eqnarray}
\label{80}
x^\mu \,
\OP^{\tw\kln{\tau_0+2j}\;\text{A/B}}_{\mu|n-2j}\kln{x} &=&
\OP^{\tw\kln{\tau_0+2j}\;\text{A/B}}_{n+1-2j}\kln{x} \; ,
\\
\label{81}
 x^\mu \,
\OP^{\tw\kln{\tau_0+1+2j}\;\text{C}}_{\mu|n-2j}\kln{x} &=& 0 \; .
\end{eqnarray}
%!p____________________________________________________________________________________________________________________________________
The first of these relations is obtained \, due to the factor
$1/\kln{n+1-2j}$, the second one due to the `internal'
antisymmetry of the vector operator (\ref{compx_twist_op3_vec}).

We have written the operators
$\OP^{\tw\kln{\tau_0+2j}\;\text{A}}_{\mu|n-2j}\kln{x}$,
$\OP^{\tw\kln{\tau_0+2j}\;\text{B}}_{\mu|n-2j}\kln{x}$ and
$\OP^{\tw\kln{\tau_0+1+2j}\;\text{C}}_{[\mu\sigma]|n-1-2j}\kln{x}$
also in the form (\ref{compx_twist_op1_vector}),
(\ref{compx_twist_op2_vector}) and (\ref{compx_twist_op3b})
because in the on-cone limit, $x^2 \ra 0$, the projector $H_{n}$
reduces to the identity.

With these definitions the expansion (\ref{compx_twist_sym1})
related to symmetry type I, i.e., totally symmetric local
operators, reads
%!p____________________________________________________________________________________________________________________________________
\begin{align}
\label{compx_twist_sym2}
 \Y^{1\;\rho}_{\mu|n} \; \OP_{\rho|n}\kln{x}
&= \frac{1}{n+1} \, \sum_{j=0}^{ \kle{ \frac{n+1}{2} } } \frac{
\kln{n+2-2j}! }{ 4^j \, j! \, \kln{n+2-j}! } \;\, \pd_\mu \,
\kln{x^2}^{j} \;
\OP^{\tw\kln{\tau_0+2j}\;\text{A}}_{n+1-2j}\kln{x}
\\
\nonumber &= \frac{1}{n+1} \, \sum_{j=0}^{ \kle{ \frac{n+1}{2} } }
\frac{ \kln{n+2-2j}! }{ 4^j \, j! \, \kln{n+2-j}! } \;\,
\kln{x^2}^{j-1} \KLE{ x^2 \kln{n+1-2j} \;
\OP^{\tw\kln{\tau_0+2j}\;\text{A}}_{\mu|n-2j}\kln{x}
 + 2j \; x_\mu \, \OP^{\tw\kln{\tau_0+2j}\;\text{A}}_{n+1-2j}\kln{x} }
 \,,
\\ %\end{eqnarray}
%!p____________________________________________________________________________________________________________________________________
\intertext{and the expansion (\ref{compx_twist_symII}), globally
related to symmetry type II, reads }
%!p____________________________________________________________________________________________________________________________________
%\begin{eqnarray}
\label{compx_twist_sym3}
 \Y^{2\;\rho}_{\mu|n} \; \OP_{\rho|n}\kln{x}
&=
 \sum_{j=0}^{ \kle{ \frac{n+1}{2} } } \frac{ \kln{n+1-2j}! }{ 4^j \, j! \, \kln{n+1-j}! } \, \kln{x^2}^{j-1} x^\sigma
\nonumber\\
\nonumber &
 \times \KLEEo{ x^2 \; \OP_{\kle{\mu\sigma}|n-1-2j}^{\tw\kln{\tau_0+1+2j} \;\text{C} } } \oKLEE{ \, - \, \frac{ 8j \kln{n+2-2j}
}{n+3-2j} \; x_{[\mu} \, \kln{ \frac{1}{n+1} \;
\OP_{\sigma]|n-2j}^{\tw\kln{\tau_0+2j} \; \text{A} } -
\OP_{\sigma]|n-2j}^{\tw\kln{\tau_0+2j} \; \text{B} } } }
\\
\intertext{}
 &=
 \sum_{j=0}^{ \kle{ \frac{n+1}{2} } } \frac{ \kln{n+1-2j}! }{ 4^j \, j! \, \kln{n+1-j}! } \, \kln{x^2}^{j-1}
\\
\nonumber &
 \times \KLEEo{ x^2 \; \OP_{\mu|n-2j}^{\tw\kln{\tau_0+1+2j} \;\text{C} } } \, + \, \frac{ 4j \kln{n+2-2j}
}{n+3-2j} \; \KLSSo{} x^2 \kln{ \frac{1}{n+1} \; \OP_{\mu|n-2j}^{\tw\kln{\tau_0+2j} \; \text{A} } -
\OP_{\mu|n-2j}^{\tw\kln{\tau_0+2j} \; \text{B} } }
\\
\nonumber & \qquad \qquad \qquad \qquad \qquad \qquad \qquad
\qquad \qquad \;\; - \; x_\mu \kln{ \frac{1}{n+1} \;
\OP_{n+1-2j}^{\tw\kln{\tau_0+2j} \; \text{A} } -
\OP_{n+1-2j}^{\tw\kln{\tau_0+2j} \; \text{B} } } \oKLSS{} \oKLEE{}
\; .
\end{align}
%!p____________________________________________________________________________________________________________________________________
Now, putting together both contributions, Eqs.
(\ref{compx_twist_sym2}) and (\ref{compx_twist_sym3}), we finally
obtain the {\em infinite twist-decomposition of the complete local
operator} as follows:
%!p____________________________________________________________________________________________________________________________________
%\begin{center}
%\fbox{
%\begin{minipage}{16cm}
\begin{eqnarray}
\label{compx_fertig1}
 \OP_{\mu|n}\kln{x}
&=&
 \sum_{j=0}^{ \kle{ \frac{n+1}{2} } } \frac{ \kln{n+1-2j}! }{ 4^j \, j! \, \kln{n+1-j}! } \, \kln{x^2}^{j-1} \KLEEo{ x^2 \,
\OP_{\mu|n-2j}^{\tw\kln{\tau_0+1+2j} \;\text{C} } }\kln{x}
\\
\nonumber
&&
 \qquad \quad + \frac{ n+2-2j }{ n+3-2j } \KLSSo{ \frac{1}{n+2-j} \KLN{ \kln{n+3} \; x^2 \, \OP_{\mu|n-2j}^{\tw\kln{\tau_0+2j} \; \text{A} }\kln{x}
- 2j \, x_\mu \; \OP_{n+1-2j}^{\tw\kln{\tau_0+2j} \; \text{A} }\kln{x} } \;
 }
\\
\nonumber
&&
 \qquad \qquad \qquad \qquad \qquad \qquad \qquad \qquad \,
\, \oKLEE{ \oKLSS{ - \; 4j \KLN{ x^2 \, \OP_{\mu|n-2j}^{\tw\kln{\tau_0+2j} \; \text{B} }\kln{x}
- x_\mu \; \OP_{n+1-2j}^{\tw\kln{\tau_0+2j} \; \text{B} }\kln{x} } \;
 } } \; .
\end{eqnarray}
%\end{minipage} }
%\end{center}
%!p____________________________________________________________________________________________________________________________________

At this stage we observe that one can add up both twist
contributions appearing behind the $x_\mu$ trace terms to a single
operator. Namely, we find an expression containing the generalized
interior derivative $\D^{\rho}$:
%!p____________________________________________________________________________________________________________________________________
\begin{eqnarray}
\label{beobachtung}
 - \,\hbox{\large$\frac{1}{2}$}\,\OP_{n+1-2j}^{\tw\kln{\tau_0+2j} \; \text{A} }\kln{x}
 +  (n+2-j) \; \OP_{n+1-2j}^{\tw\kln{\tau_0+2j} \; \text{B}
 }\kln{x} =  H_{n+1-2j} \; \D^{\rho} \;
 \Box^{j-1} \, O_{\rho|n}\kln{x} \; .
\end{eqnarray}
%!p____________________________________________________________________________________________________________________________________
However, since the vector operators of type A and B cannot be
combined in similarly this may just be a coincidence. Furthermore,
vector operators of type B do not appear in the on-cone limit
while vector operators of type A are present in this case. One can
therefore not expect to find a useful combination of these
operators in terms of $\D^{\rho}$. Of course, this justifies the
distinction of operators into type A and type B. The open question
therefore is whether relation (\ref{beobachtung}) is genuine for
general trace terms surviving the on-cone limit or not. To judge
this question we would need the full off-cone twist decomposition
of a general 2nd rank tensor operator which is not available by
now.

The result (\ref{compx_fertig1}) is valid for a generic  local
vector operator $\OP_{\mu|n}\kln{x}$ and we are now going to apply
it to the local operators $O_{\mu|n}\kln{x}$ and
$M_{\mu|n}\kln{x}$ which, then, will be summed up to the non-local
operators. Since the `internal' structure of these operators is
different (compare the moments (\ref{compx_momo_ex}) and
(\ref{compx_momm_ex})) we will now treat these operators
separately.

%%%%%%%%%%%%%%%%%%%%%%%%%%%%%%%%%%%%%%%%%%%%%%%%%%%%%%%%%%%%%%%%%%%
%%%%%%%%%%%%%%%%%%%%%%%%%%%%%%%%%%%%%%%%%%%%%%%%%%%%%%%%%%%%%%%%%%%
\section{Complete twist-decomposition of the nonlocal operators}
%for $O_\mu\kln{\kappa\,x,-\kappa\,x}$}
%%%%%%%%%%%%%%%%%%%%%%%%%%%%%%%%%%%%%%%%%%%%%%%%%%%%%%%%%%%%%%%%%%%
%%%%%%%%%%%%%%%%%%%%%%%%%%%%%%%%%%%%%%%%%%%%%%%%%%%%%%%%%%%%%%%%%%%

This chapter is devoted to the infinite twist decomposition of the
nonlocal operators $O_\mu\kln{\kappa\,x,-\kappa\,x}$ and
$M_\mu\kln{\kappa\,x,-\kappa\,x}$, Eqs.~(\ref{vector_op}) and
(\ref{def_Mmu}), by summing up their corresponding local
decomposition. Also here we suppress anti-symmetrization w.r.t.
$\kappa$ which can be done at the end. Obviously, for the operator
$O_\mu\kln{\kappa\,x,-\kappa\,x}$ the above expansion
(\ref{compx_fertig1}) is already the complete twist-decomposition
of $O_{\mu|n}\kln{x}$ and will not be repeated for that reason.
However, for the operator $M_\mu\kln{\kappa\,x,-\kappa\,x}$ the
expansion (\ref{compx_fertig1}) reduces to operators of type B and
C only, i.e., operators of type A which are related to symmetry
type I have to be omitted.

When the explicit expressions for the moments $O_{\mu|n}\kln{x}$
and $M_{\mu|n+1}\kln{x}$ given by (\ref{compx_momo_ex}) and
(\ref{compx_momm_ex}), respectively, are put into the local
operators (\ref{compx_twist_op1}) -- (\ref{compx_twist_op3}) these
operators of well-defined twist may be expressed in terms of
Gegenbauer polynomials.
For this reason let us introduce some useful definitions and
relations. First, a set of polynomials $h^\nu_{n}\kln{u|x}$, being
symmetric and homogeneous in the variables $u$ and $x$, is defined
as follows:
\begin{alignat}{4}
\label{compx_hnu}
 h_n^\nu\left( u|x \right) &:\;= \left(
\hbox{\large$\frac{1}{2}$}\sqrt{u^2 x^2} \right)^n \;
C_n^{\nu}\left( \frac{ux}{\sqrt{u^2 x^2}} \right) & \quad
\text{for} \quad & n \geq 0
\\
\nonumber h_n^\nu\left( u|x \right) & :\;= 0 &
\quad\text{for}\quad & n < 0\;,
\end{alignat}
where $C^\nu_n\left( z \right)$ are the Gegenbauer polynomials
(see Ref.~\cite{PBM}, Appendix II.11) given by
%!p____________________________________________________________________________________________________________________________________
\begin{eqnarray}
\label{compx_gegenbauer}
 C_n^\nu\kln{z}
&=&
 \frac{1}{\kln{\nu-1}!} \, \sum_{k=0}^{\kle{\frac{n}{2}}} \, \frac{\kln{-1}^k\,\kln{n-k+\nu-1}!}{k!\,\kln{n-2k}!} \; \kln{2\,z}^{n-2k}
\qquad \text{for} \qquad \nu > 0 \; ,
\\
\label{compx_gegenbauer_null}
 C_n^0\kln{z}
&=&
 \sum_{k=0}^{\kle{\frac{n}{2}}} \,
 \frac{\kln{-1}^k\,\kln{n-k-1}!}{k!\,\kln{n-2k}!} \; \kln{2\,z}^{n-2k}
 \qquad \text{for}\qquad n> 0\,,  \qquad C_0^0\kln{z} = 1\; .
\end{eqnarray}
%!p____________________________________________________________________________________________________________________________________
For $\nu > 0$, the polynomials $h_n^\nu\kln{u|x}$ obey the following relations,
%!p____________________________________________________________________________________________________________________________________
\begin{eqnarray}
\label{compx_rela_gegen1}
 H_n \, \kln{ux}^n
&=&
 h_n^1\kln{u|x} \; ,
\\
\label{compx_rela_gegen2}
 \pd_\alpha^x \; h_n^\nu\kln{u|x}
&=&
 {\nu} \KLn{  \, u_\alpha \, h_{n-1}^{\nu+1}\kln{u|x} -
 \hbox{\large$\frac{1}{2}$} u^2 \, x_\alpha \, h_{n-2}^{\nu+1}\kln{u|x} } \; ,
\\
\label{compx_rela_gegen3}
 \pd_\alpha^u \; h_n^\nu\kln{u|x}
&=&
 {\nu} \KLn{  \, x_\alpha \, h_{n-1}^{\nu+1}\kln{u|x} -
 \hbox{\large$\frac{1}{2}$}x^2 \, u_\alpha \, h_{n-2}^{\nu+1}\kln{u|x} } \; ,
\\
\label{compx_rela_gegen4}
 \Box \, h_n^\nu\kln{u|x}
&=&
 \nu\kln{\nu-1} \, u^2 \, h_{n-2}^{\nu+1}\kln{u|x} \; ,
\\
\label{compx_rela_gegen5}
 \Box_u \, h_n^\nu\kln{u|x}
&=&
 \nu\kln{\nu-1} \, x^2 \, h_{n-2}^{\nu+1}\kln{u|x} \; ,
\\
\label{compx_rela_gegen6}
 \kln{n+\nu} \, h_n^\nu\kln{u|x}
&=&
 {\nu} \KLn{  \, h_n^{\nu+1}\kln{u|x} -
 \hbox{\large$\frac{1}{4}$} u^2 x^2 \; h_{n-2}^{\nu+1}\kln{u|x} } \; ,
\end{eqnarray}
%!p____________________________________________________________________________________________________________________________________
which can be traced back to the definition and corresponding
relations of the Gegenbauer polynomials. %(\ref{compx_gegenbauer})
%or to similar relations obeyed by the $C_n^\nu$, see reference \cite{PBM} page 737.
They will be used to derive the representations
(\ref{compx_twist_o_op1_b}) -- (\ref{compx_twist_o_op3_b}) below.
Relation (\ref{compx_rela_gegen1}) is the origin for all
Gegenbauer polynomials appearing in these expressions.

%%%%%%%%%%%%%%%%%%%%%%%%%%%%%%%%%%%%%%%%%%%%%%%%%%%%%%%%%%%%%%%%%%%%%%%
\subsection{Infinite twist-decomposition of the operator
 $O_\mu\kln{\kappa\,x,-\kappa\,x}$}
%%%%%%%%%%%%%%%%%%%%%%%%%%%%%%%%%%%%%%%%%%%%%%%%%%%%%%%%%%%%%%%%%%%%%%%%%

Now, as announced, let us rewrite the operators of definite twist
related to $O_\mu\kln{\kappa\,x,-\kappa\,x}$ according to
(\ref{compx_twist_op1}) -- (\ref{compx_twist_op3}) in terms of the
polynomials (\ref{compx_hnu}):
\\
For the two scalar operators we obtain:
%!p____________________________________________________________________________________________________________________________________
\begin{align}
%\label{compx_twist_o_op1}
\nonumber
 O^{\tw\kln{2+2j}\;\text{A}}_{n+1-2j}\kln{x}
%&=&
% H_{n+1-2j} \; \Box^j \; x^\rho \; O_{\rho|n}\kln{x}
%\\
%\nonumber
%&=
% \frac{n!}{\kln{n+1-2j}!} \; \int \text{d}^4\!u \; O_{\rho}\kln{u} \, \pd_u^\rho \; \kln{u^2}^{j} \; h^1_{n+1-2j}\kln{u|x}
%\\ \nonumber
 &=
 \frac{n!}{2\,\kln{n+1-2j}!} \; \int \text{d}^4\!u \; O_{\rho}\kln{u} \, \kln{u^2}^{j-1}
\\
\label{compx_twist_o_op1_b}
 &
 \qquad \times \KLE{ 4\, j \, u^\rho \; h^1_{n+1-2j}\kln{u|x} + \, u^2 \KLn{ 2\,x^\rho \; h^2_{n-2j}\kln{u|x} - x^2 \, u^\rho \;
h^2_{n-1-2j}\kln{u|x} } } \; ,
\\
%\label{compx_twist_o_op2}
%\nonumber
 O^{\tw\kln{2+2j}\;\text{B}}_{n+1-2j}\kln{x}
%&=&
% H_{n+1-2j} \; \Box^{j-1} \; \pd^\rho \; O_{\rho|n}\kln{x}
%\\
\label{compx_twist_o_op2_b}
 &=
 \frac{n!}{\kln{n+1-2j}!} \; \int \text{d}^4\!u \; O_{\rho}\kln{u} \, \kln{u^2}^{j-1} u^\rho \; h^1_{n+1-2j}\kln{u|x} \; .
\\ %\end{eqnarray}
%!p____________________________________________________________________________________________________________________________________
\intertext{The two vector operators
(\ref{compx_twist_op1_vector_def}) and
(\ref{compx_twist_op2_vector_def}) are obtained from the previous
expressions as follows:}
%!p____________________________________________________________________________________________________________________________________
%\begin{eqnarray}
%\label{compx_twist_o_op1_vec}
 O^{\tw\kln{2+2j}\;\text{A}}_{\mu|n-2j}\kln{x}
%&=&
% \frac{1}{n+1-2j} \; \pd_\mu \; H_{n+1-2j} \; \Box^j \; x^\rho \; O_{\rho|n}\kln{x}
%\\
\nonumber
 &=
 \frac{n!}{2\,\kln{n+1-2j}\kln{n+1-2j}!} \; \int \text{d}^4\!u \; O_{\rho}\kln{u} \, \kln{u^2}^{j-1}
\\
\label{compx_twist_o_op1_b_vec}
 &
 \qquad \times \; \pd_\mu \KLE{ 4\, j \, u^\rho \; h^1_{n+1-2j}\kln{u|x} + \, u^2 \KLn{ 2\,x^\rho \; h^2_{n-2j}\kln{u|x} - x^2 \, u^\rho \;
h^2_{n-1-2j}\kln{u|x} } } \; ,
\\
%\label{compx_twist_o_op2_vec}
 O^{\tw\kln{2+2j}\;\text{B}}_{\mu|n-2j}\kln{x}
% &=
% \frac{1}{n+1-2j} \; \pd_\mu \; H_{n+1-2j} \; \Box^{j-1} \; \pd^\rho \; O_{\rho|n}\kln{x}
%\\
\label{compx_twist_o_op2_b_vec}
 &=
 \frac{n!}{\kln{n+1-2j}\kln{n+1-2j}!} \; \int \text{d}^4\!u \; O_{\rho}\kln{u} \, \kln{u^2}^{j-1} u^\rho \; \pd_\mu \; h^1_{n+1-2j}\kln{u|x} \; .
\\ %\end{eqnarray}
%!p____________________________________________________________________________________________________________________________________
\intertext{The vector operator of type C is given by}
%!p____________________________________________________________________________________________________________________________________
%\begin{eqnarray}
%\label{compx_twist_o_op3} %\hspace{-.8cm}
 O^{\tw\kln{3+2j}\;\text{C}}_{\mu|n-2j}\kln{x}
%&=
% \frac{2}{n+1-2j} \; x^\sigma \kln{ \delta_{[\mu}^\rho \, \pd_{\sigma]} + \frac{1}{n+1-2j} \; \X_{[\mu\sigma]} \; \pd^{\rho} } \; H_{n-2j} \;
%\Box^{j} \; O_{\rho|n}\kln{x}
%\\
\label{compx_twist_o_op3_b} %\hspace{-.8cm}
 &=
 \frac{2\;n!\;\;x^\sigma}{\kln{n+1-2j}\kln{n+1-2j}!} \kln{ \delta_{[\mu}^\rho \pd_{\sigma]} \kln{n+1-2j}
 + \X_{[\mu\sigma]} \, \pd^\rho } \! \int \text{d}^4\!u \; O_{\rho}\kln{u} \, \kln{u^2}^{j} h^1_{n-2j}\kln{u|x} .
\end{align}
%!p____________________________________________________________________________________________________________________________________

In order to have a cross-check on our result (\ref{compx_fertig1})
we perform the on-cone limit $x \ra \tx,\,\tx^2= 0$, for
$O_{\mu|n}\kln{x}$ and find
%!p____________________________________________________________________________________________________________________________________
\begin{equation}
 O_{\mu|n}\kln{\tx} = O_{\mu|n}^{\tw2 \; \text{A} }\kln{\tx} +
O_{\mu|n}^{\tw3 \; \text{C} }\kln{\tx} -
\frac{\tx_\mu}{2\kln{n+1}^2} \; \KLn{O_{n-1}^{\tw4 \; \text{A}
}\kln{\tx}-  2(n+1) O_{n-1}^{\tw4 \; \text{B} }\kln{\tx}}
\end{equation}
%!p____________________________________________________________________________________________________________________________________
with
%!p____________________________________________________________________________________________________________________________________
\begin{eqnarray}
% O_{\mu|n}^{\tw2 \; \text{A} }\kln{\tx} &=&
% \frac{1}{\kln{n+1}^2} \; \D_\mu \;  x^\rho \, O_{\rho|n}\kln{\tx} \; ,
%\\  O_{\mu|n}^{\tw3 \; \text{C} }\kln{\tx} &=&
% \frac{2}{n\kln{n+1}} \; x^\sigma \kln{ \delta_{[\mu}^\rho \, \D_{\sigma]}
%+ \frac{1}{n+1} \; \X_{[\mu\sigma]} \; \D^{\rho} } O_{\rho|n}\kln{\tx} \;,
\label{compx_summe_o1}
 O_{\mu|n}^{\tw2 \; \text{A} }\kln{\tx}
&=&
 \frac{1}{n+1}\kln{ \pd_\mu
 - \frac{1}{2\kln{n+1}} \; x_\mu \; \Box } \, x^\rho \, O_{\rho|n}\kln{x} \PIpe{x=\tx} \; ,
\\
\label{compx_summe_o2}
 O_{\mu|n}^{\tw3 \; \text{C} }\kln{\tx}
&=&
 \frac{1}{n+1} \kln{ n \, \delta_{\mu}^{\rho}
 - x^\rho \, \pd_\mu - \frac{x_\mu}{n+1} \KLn{ \kln{n-1} \, \pd^\rho - x^\rho \, \Box }
} O_{\rho|n}\kln{x} \PIpe{x=\tx} \; ,
\\
 O_{n-1}^{\tw4 \; \text{A} }\kln{\tx}
&=&
 \Box \; x^\rho \, O_{\rho|n}\kln{\tx} \; ,
\\
 O_{n-1}^{\tw4 \; \text{B} }\kln{\tx}
&=&
 \pd^\rho \, O_{\rho|n}\kln{\tx} \; .
\end{eqnarray}
%!p____________________________________________________________________________________________________________________________________
Thus, only the twist-2, the twist-3 and two twist-4 parts survive.
The twist-4 parts are combined according to
%!p____________________________________________________________________________________________________________________________________
\begin{eqnarray}
\label{compx_summe_o3_b}
 - \frac{1}{2} %\kln{n+1}^2}
  O_{n-1}^{\tw4 \; \text{A} }\kln{\tx}
 + (n+1) \; O_{n-1}^{\tw4 \; \text{B} }\kln{\tx}
&=&
 %\frac{1}{\kln{n+1}^2} \;
 \kln{ \kln{n+1} \, \pd^\rho -
 \frac{1}{2} \; \Box \, x^\rho } \, O_{\rho|n}\kln{x} \PIpe{x=\tx} \;
%\\
%\label{compx_summe_o3_b}
%&=&
 %\frac{1}{\kln{n+1}^2} \;
= \D^{\rho} \, O_{\rho|n}\kln{x} \PIpe{x=\tx} \; ,
\end{eqnarray}
%!p____________________________________________________________________________________________________________________________________
which can be directly deduced from (\ref{beobachtung}). Comparing
these results with those of Ref.~\cite{GL01} we find that the
twist-2, the twist-3 and the twist-4 parts obtain the form (2.22),
(2.23) and (2.24), respectively. If we perform the summation to a
non-local operator we can compare with Ref.~\cite{GLR99}. The
twist-2, twist-3 and twist-4 parts are then given by (3.26),
(3.42) and (3.45), respectively. The expansion
(\ref{compx_fertig1}) therefore provides us with the correct
on-cone limit.

In order to get the infinite twist decomposition of the non-local
operator $O_\mu\kln{\kappa \, x , - \kappa \, x}$ we perform the
summation over $n$ also off-cone. To this purpose let us introduce
an integral representation for the coefficient $c\kln{j,n}$ by
using the integral representation of Euler's beta function,
%!p____________________________________________________________________________________________________________________________________
\begin{equation}
\label{compx_euler_beta}
 \frac{ \kln{n+1-2j}! }{ \kln{n+1-j}! } = \frac{1}{ \kln{j-1}! } \; \frac{\Gamma\kln{n+2-2j}\Gamma\kln{j}}{\Gamma\kln{n+2-j}} =
\frac{1}{ \kln{j-1}! } \; \int_0^1 \text{d}t \;\, \kln{1-t}^{j-1}
\;t^{n+1-2j} ,
\end{equation}
%!p____________________________________________________________________________________________________________________________________
which is valid for $j\ge 1$. In addition, we introduce also
integral representations for the different fractions in the second
term of the expansion (\ref{compx_fertig1}),
%!p____________________________________________________________________________________________________________________________________
\begin{eqnarray}
\label{compx_summe_fraction1}
 \frac{1}{\kln{n+2-j}\kln{n+3-2j}}
&=&
 %\lim_{k\ra j} \;
 \int_0^1 \text{d} \lambda \; \frac{\kln{ 1 - \lambda^{j-1}}}{j-1} \; \lambda^{n+2-2j} \; ,
\\
\label{compx_summe_fraction2}
 \frac{1}{n+3-2j}
&=&
 \int_0^1 \text{d} \lambda \;\; \lambda^{n+2-2j} \; .
\end{eqnarray}
%!p____________________________________________________________________________________________________________________________________
These representations hold for $j\ge 0$, since the first
representation is correct also in the limit $j=1$ with the result
$-\int_0^1\text{d}\lambda\;\ln\lambda\;\lambda^n\;$.

In the same manner as in Ref.~\cite{GLR01}, the polynomials
$h_n^\nu\kln{u|x}$ which are defined in terms of Gegenbauer
polynomials may be summed up to the functions $\HB^\nu\kln{u|x}$
being defined in terms of Bessel functions,
%!p____________________________________________________________________________________________________________________________________
\begin{equation}
\label{compx_bessel}
 \HB^\nu\kln{u|x} := \sqrt{\pi} \kln{\sqrt{ \kln{ux}^2 - u^2 x^2 }}^{1/2-\nu} J_{\nu-1/2}\kln{ \frac{1}{2} \sqrt{ \kln{ux}^2 - u^2 x^2
} } \, \e^{\im \, \kln{ux}/2} \; .
\end{equation}
%!p____________________________________________________________________________________________________________________________________
 Namely, with that
definition taking into account Eq. II.5.13.1.3 of Ref.~\cite{PBM}
(or formula (2.29) from Ref.~\cite{GLR01} ) we can deduce the
formula
%!p____________________________________________________________________________________________________________________________________
\begin{eqnarray}
\label{compx_summe}
 \sum_{n=0}^\infty \frac{\im^n}{\kln{n+l}!} \; \kappa^{n-m} \;  h^\nu_{n-m}\kln{u|x}
&=&
 \frac{\im^m}{\kln{\nu-1}!} \; \kle{ \;\; \prod_{k=1+l+m}^{2\,\nu-1} \!\!\!\! \kln{ k + x\pd } } \HB^\nu\kln{u|\kappa\,x}
\\
\nonumber
&&
 \text{for} \qquad m \in \kls{0,\dots,2\,\nu-1}
 \qquad \text{and} \qquad l \in \kls{-m,\dots,2\,\nu-m-1}\,;
\end{eqnarray}
%!p____________________________________________________________________________________________________________________________________
for $k=2\nu$ the product in (\ref{compx_summe}) is empty. The
ranges for $l$ and $m$ have been chosen such that integral
representations for factors like $1/\kln{n+i}$ are avoided. In
principle these factors could be included in a more general
formula which, however, will not be needed.

Now we define non-local operators of well-defined twist by
%!p____________________________________________________________________________________________________________________________________
\begin{eqnarray}
\label{compx_summe_nla}
 O^{\tw\kln{2+2j}\;\text{A/B}}\kln{\kappa\,x,-\kappa\,x}
&:=&
 \sum_{n=0}^\infty \; \frac{\im^n}{n!} \; \kappa^{n+1-2j}
 \; O^{\tw\kln{2+2j}\;\text{A/B}}_{n+1-2j}\kln{x} \; ,
\\
\label{compx_summe_nla_vec}
 O_\mu^{\tw\kln{2+2j}\;\text{A/B/C}}\kln{\kappa\,x,-\kappa\,x}
&:=&
 \sum_{n=0}^\infty \; \frac{\im^n}{n!} \; \kappa^{n-2j}
 \; O^{\tw\kln{2+2j}\;\text{A/B/C}}_{\mu|n-2j}\kln{x} \; ,
\end{eqnarray}
%!p____________________________________________________________________________________________________________________________________
which are introduced as functions of $\kappa x$ because they will
be used in the twist decomposition of a non-local vector operator
which obeys that property!
%We have introduced factors of
%$\kappa^{n+1-2j}$ and $\kappa^{n-2j}$ in the sums
%(\ref{compx_summe_nla}) and (\ref{compx_summe_nla_vec}) to ensure
%that every $x$ appearing in the considered operators is multiplied
%by $\kappa$ and every $\pd_x$ by $1/\kappa$.
This will be very important for the corresponding summations.

Performing the above sums, using equation (\ref{compx_summe}), and
the integral representation $1/\kln{n+1-2j} = \int_0^1
\text{d}\lambda \, \lambda^{n-2j}$ we find by  straightforward
calculations,
%!p____________________________________________________________________________________________________________________________________
\begin{eqnarray}
\nonumber
 O^{\tw\kln{2+2j}\;\text{A}}\kln{\kappa\,x,-\kappa\,x}
&=&
 \frac{1}{2} \, \int \text{d}^4\!u \; O_\rho\kln{u}
 \; \kln{-u^2}^{j-1} \KLEEo{} 4 \, j \;\, \im u^\rho \, \kln{ 1 + x\pd } \;
\HB^1\kln{u|\kappa \, x}
\\
\label{compx_summea} &&
 \qquad \qquad - \, u^2 \KLn{ 2 \, \kappa \, x^\rho
 \, \kln{2+x\pd} - \im u^\rho \; \kappa^2 \, x^2 } \kln{3+x\pd} \;
\HB^2\kln{u|\kappa \, x} \oKLEE{} \; ,
\\
\label{compx_summeb}
 O^{\tw\kln{2+2j}\;\text{B}}\kln{\kappa\,x,-\kappa\,x}
&=&
 \int \text{d}^4\!u \; O_\rho\kln{u} \; \kln{-u^2}^{j-1}
 \, \im u^\rho \, \kln{1+x\pd} \; \HB^1\kln{u|\kappa \, x} \; ,
 \qquad\qquad\qquad j \ge 1\,,
\\
\nonumber
 O_\mu^{\tw\kln{2+2j}\;\text{A}}\kln{\kappa\,x,-\kappa\,x}
&=&
 \frac{1}{2\,\kappa} \, \int \text{d}^4\!u \; O_\rho\kln{u}
 \; \kln{-u^2}^{j-1} \; \pd_\mu \; \int_0^1 \frac{\text{d}\lambda}{\lambda}
\KLEEo{} 4 j \; \im u^\rho \, \kln{ 1 + x\pd } \;
\HB^1\kln{u| \lambda \kappa \, x}
\\
\label{compx_summea_vec}
&&
 \qquad  - \, u^2 \KLn{ 2 \, \lambda\kappa \; x^\rho
 \, \kln{2+x\pd} - \im u^\rho \kln{\lambda \kappa}^2 \, x^2 } \kln{3+x\pd} \;
\HB^2\kln{u| \lambda \kappa \, x} \oKLEE{} \; ,
\\
\label{compx_summeb_vec}
 O_\mu^{\tw\kln{2+2j}\;\text{B}}\kln{\kappa\,x,-\kappa\,x}
&=&
 \frac{1}{\kappa} \int \text{d}^4\!u \; O_\rho\kln{u}
 \; \kln{-u^2}^{j-1} \, \im u^\rho \; \pd_\mu \; \kln{1+x\pd} \int_0^1 \frac{\text{d}\lambda}{\lambda} \;
 \HB^1\kln{u| \lambda \kappa \, x} \,, \qquad j \ge 1\,,
\\
\label{compx_summec}
 O^{\tw\kln{3+2j}\;\text{C}}_{\mu}\kln{\kappa\,x,-\kappa\,x}
&=&
 2 \int \text{d}^4\!u \; O_\rho\kln{u} \; \kln{-u^2}^j
 \; x^\sigma \kln{ \delta_{[\mu}^\rho \pd_{\sigma]} \kln{ 1 + x\pd } +
\X_{[\mu\sigma]} \, \pd^\rho } \int_0^1 \text{d} \lambda
\;\, \HB^1\kln{u|\lambda \kappa \, x} \; .
\end{eqnarray}
%!p____________________________________________________________________________________________________________________________________
In the case $j=0$ we can again compare with Ref.~\cite{GLR01}: The
twist-2 operator $O^{\tw2\;\text{A}}\kln{\kappa\,x,-\kappa\,x}$ is
then equal to (3.15) and the twist-3 operator $x^\sigma
O^{\tw3\;\text{C}}_{[\mu\sigma]}\kln{\kappa\,x,-\kappa\,x}$ is
equal to (4.27). Operators of type B, (\ref{compx_summeb}) and
(\ref{compx_summeb_vec}), have not been considered in \cite{GLR01}
because the full decomposition of operators related to symmetry
type II was not available there.

Now we are in a position to sum up to a non-local operator using
equations (\ref{compx_euler_beta}) and
(\ref{compx_summe_fraction1}) which are valid for $j\ge 1$.
%!p____________________________________________________________________________________________________________________________________
%\begin{eqnarray}\label{compx_summe1}
% \frac{1}{n+1}&=& \int_0^1 \text{d}\lambda \; \lambda^n \; ,
%\\ \label{compx_summe2} \frac{1}{\kln{n+1}^2}
%&=& - \int_0^1 \text{d}\lambda \; \ln\lambda \; \lambda^n \; .
%\end{eqnarray}
%!p____________________________________________________________________________________________________________________________________
 The final result for the {\em infinite twist-decomposition of the
non-local operator} $O_\mu\kln{\kappa \,x,-\kappa\,x}$ obtains the
following form:
%!p____________________________________________________________________________________________________________________________________
%\begin{center}
%\fbox{
%\begin{minipage}[c]{17cm}
\begin{eqnarray}
\label{compx_fertig_nl1} \hspace{-1cm}
 &&
 O_{\mu}\kln{\kappa \, x, - \kappa \, x}
%\\ \nonumber &&
 = O_\mu^{\tw2\;\text{A}}\kln{\kappa\,x,-\kappa\,x}
\; + \; O^{\tw3\;\text{C}}_{\mu}\kln{\kappa\,x,-\kappa\,x}
\\
\nonumber \hspace{-1cm}
 &&
\qquad \qquad\qquad \quad + \sum_{j=1}^{\infty} \; %\lim_{k\ra j} \;
 \frac{\kln{\kappa^2 x^2}^{j-1}}{ 4^j \, j! \, \kln{j-1}! } \;
 \int_0^1 \frac{\text{d}t}{t} \; \kln{1-t}^{j-1}
\\
\nonumber \hspace{-1cm} &&
 \qquad \qquad\qquad \quad \quad
 \times\,\KLEEo{ } \kln{x\pd} \int_0^1 \text{d}\lambda \; \KLSSo{}
 \frac{1-\lambda^{j-1}}{j-1} \KLNo{}
\kln{1+2j+x\pd} \; \kln{\lambda t \kappa}^2 x^2 \,
O_\mu^{\tw\kln{2+2j} \; \text{A} }\kln{\lambda t\kappa\,x,-\lambda
t\kappa\,x}
\\
\nonumber \hspace{-1cm} &&
 \qquad\qquad\qquad \qquad \qquad \qquad  \qquad \qquad \qquad \qquad \qquad \qquad
 - \; 2j \, \kln{ \lambda t \kappa } \; x_\mu
\; O^{\tw\kln{2+2j} \; \text{A} }\kln{\lambda t\kappa\,x,-\lambda
t\kappa\,x} \oKLN{}
\\
\nonumber \hspace{-1cm} &&
\qquad \qquad \qquad\qquad \qquad\quad  %\qquad
- \; 4j \, \KLN{ \kln{\lambda t \kappa}^2 x^2 \,
  O_\mu^{\tw\kln{2+2j} \; \text{B} }\kln{\lambda t\kappa\,x,-\lambda t\kappa\,x}
-  \, \kln{ \lambda t \kappa } \; x_\mu \; O^{\tw\kln{2+2j} \;
\text{B} }\kln{\lambda t\kappa\,x,-\lambda t\kappa\,x} } \oKLSS{}
\\
\nonumber \hspace{-1cm} &&
 \qquad \qquad \qquad\qquad +
 \; \kln{t\kappa}^2 x^2 \; O^{\tw\kln{3+2j}\;\text{C}}_{\mu}\kln{t\kappa\,x,-t\kappa\,x}
\oKLEE{} \; .
\end{eqnarray}
%\end{minipage} }
%\end{center}
Thereby, $ \lim_{j\ra 1} \; \left((1-\lambda^{j-1})/(j-1)\right) =
- \ln\lambda $ has to be observed.

The above formula shows explicitly the origin of the different
twist contributions. The twist-2 and twist-3 contributions are
given by single terms which are traceless but twist-4 and also all
higher contributions of even twist consist of two terms where the
traceless operators
$O_\mu^{\tw4\;\text{A/B}}\kln{\lambda\kappa\,x,-\lambda\kappa\,x}$
and $O^{\tw4\;\text{A/B}}\kln{\lambda\kappa\,x,-\lambda\kappa\,x}$
appear behind trace terms of type $x^2$ and $x_\mu$. Therefore,
the sum of these two twist-4 contributions can never be traceless.
Contracting the expression (\ref{compx_fertig_nl1}) with $x^\mu$
we obtain after a straightforward computation the well-known infinite twist
decomposition of $O\kln{\kappa\,x,-\kappa\,x}$, due to
Eqs.~(\ref{80}) and (\ref{81}):
%Contracting the expression (\ref{compx_fertig_nl1}) with $x^\mu$
%after a straightforward computation the well-known infinite twist
%decomposition of $O\kln{\kappa\,x,-\kappa\,x}$ obtains, due to
%Eqs.~(\ref{80}) and (\ref{81}):
\begin{eqnarray}
\label{compx_fertig_nl10}
  O\kln{\kappa \, x, - \kappa \, x}
%\\ \nonumber &&
 &=& O^{\tw2\;\text{A}}\kln{\kappa\,x,-\kappa\,x}
%\\ \nonumber \hspace{-1cm}  && \qquad \qquad\qquad \quad
+ \sum_{j=1}^{\infty} \; %\lim_{k\ra j} \;
 \frac{\kln{\kappa^2 x^2}^{j-1}}{ 4^j \, j! \, \kln{j-1}! } \;
 \kln{x\pd} \,\kln{1+x\pd}\int_0^1 \frac{\text{d}t}{t} \; \kln{1-t}^{j-1}
\\ \nonumber  &&
 \qquad \qquad\qquad \qquad \qquad\qquad %\qquad \qquad
 \times\,
  \int_0^1 \text{d}\lambda \; %\KLSSo{}
 \frac{1-\lambda^{j-1}}{j-1} %\KLNo{}
 \; \kln{\lambda t \kappa}^2 x^2 \,
O^{\tw\kln{2+2j} \; \text{A} }\kln{\lambda t\kappa\,x,-\lambda
t\kappa\,x} %\oKLN{}
 \; ,
\end{eqnarray}
which already has been determined in Ref.~\cite{GLR01},
Eqs.~(3.14,15).

It is obvious that the contributions of even twist, beginning with
twist-4, look rather complicated. In principle, it is possible to
express the double integral $\int_0^1 \text{d} t \, \int_0^1
\text{d} \lambda$ in Eqs.~(\ref{compx_fertig_nl1}) and
(\ref{compx_fertig_nl10}) by a single integral representation but
then it yields combinations of some generalized hypergeometric
functions $F_{[p,q]}\Kln{[a_1,\dots,a_p],[b_1,\dots,b_q],z}$. We
did not make use of them.

%%%%%%%%%%%%%%%%%%%%%%%%%%%%%%%%%%%%%%%%%%%%%%%%%%%%%%%%%%%%%%%%%%%%%%%%%%%%%%%%%%%%%%%%%%%%%%%%%%%%%%%%%%%%%%%%%%%%%%%%%%%%%%%%%%%%%%
\subsection{Infinite twist-decomposition of $M_\mu\kln{\kappa\,x,-\kappa\,x}$}
%%%%%%%%%%%%%%%%%%%%%%%%%%%%%%%%%%%%%%%%%%%%%%%%%%%%%%%%%%%%%%%%%%%%%%%%%%%%%%%%%%%%%%%%%%%%%%%%%%%%%%%%%%%%%%%%%%%%%%%%%%%%%%%%%%%%%%

In this subsection we apply the result (\ref{compx_fertig1}) to
the operator $M_\mu\kln{\kappa\,x,-\kappa\,x}$. Since operators of
type A, (\ref{compx_twist_op1}) and
(\ref{compx_twist_op1_vector_def}), are not present in this case
the infinite twist-decomposition for the moments
$M_{\mu|n+1}\kln{x}$ obtains a simpler form:
%!p____________________________________________________________________________________________________________________________________
\begin{eqnarray}
\label{compx_twist_M}
 M_{\mu|n+1}\kln{x}
&=&
 \sum_{j=0}^{ \kle{ \frac{n+2}{2} } } \frac{ \kln{n+2-2j}! }{ 4^j \, j! \, \kln{n+2-j}! } \, \kln{x^2}^{j-1} x^\sigma
\\
\nonumber
&&
 \times \KLEEo{ x^2 \; M_{\kle{\mu\sigma}|n-2j}^{\tw\kln{2+2j}\;\text{C}}\kln{x} } \oKLEE{ \, + \, \frac{ 8\,j \kln{n+3-2j}
}{n+4-2j} \; x_{[\mu} \; M_{\sigma]|n+1-2j}^{\tw\kln{1+2j} \; \text{B} }\kln{x} }
\\
\nonumber
&=&
 \sum_{j=0}^{ \kle{ \frac{n+2}{2} } } \frac{ \kln{n+2-2j}! }{ 4^j \, j! \, \kln{n+2-j}! } \, \kln{x^2}^{j-1}
\\
\nonumber
&&
 \times \KLEEo{ x^2 \; M_{\mu|n+1-2j}^{\tw\kln{2+2j}\;\text{C}}\kln{x} } \oKLEE{ \, - \, \frac{ 4\,j \kln{n+3-2j}
}{n+4-2j} \kln{ x^2 \; M_{\mu|n+1-2j}^{\tw\kln{1+2j} \; \text{B}
}\kln{x} - \, x_{\mu} \; M_{n+2-2j}^{\tw\kln{1+2j} \; \text{B}
}\kln{x} } }\,,
\end{eqnarray}
%!p____________________________________________________________________________________________________________________________________
globally being of symmetry type II. Of course, the operators
$M_{\mu|n+1-2j}^{\tw\kln{1+2j} \; \text{B} }\kln{x}$ and
$M_{n+2-2j}^{\tw\kln{1+2j} \; \text{B} }\kln{x}$ are related to
symmetry type I. Furthermore, notice that one has to perform a
global shift $n\ra n+1$ in order to get the correct order %$n+1$
of the operator.

In principle this result follows from the complete off-cone
decomposition of the skew tensor operator $M_{[\mu\nu]|n}\kln{x}$
(see the definition (\ref{def_Mmu})) which contains contributions
of symmetry type II and III. Type~III contributions cancel
completely once they are contracted with $x^\nu$ while the type~II
contributions yield additional trace terms which are contained in
(\ref{compx_twist_M}) after the contraction. The decomposition
(\ref{compx_twist_M}) is therefore not equivalent to the type~II
part of the antisymmetric tensor case.

If we insert the explicit form of the moments $M_{\mu|n+1}\kln{x}$
given by (\ref{compx_momm_ex}) into (\ref{compx_twist_op2}),
(\ref{compx_twist_op2_vector_def})  and (\ref{compx_twist_op3})
and use (\ref{compx_rela_gegen1}) -- (\ref{compx_rela_gegen3}) we
find the following three operators of well-defined twist, again
expressed in terms of Gegenbauer polynomials,
%!p____________________________________________________________________________________________________________________________________
\begin{eqnarray}
\hspace{-.5cm} \label{compx_twist_m_op2}
 M^{\tw\kln{1+2j}\;\text{B}}_{n+2-2j}\kln{x}
%&=&
% H_{n+2-2j} \; \Box^{j-1} \; \pd^\rho \; M_{\rho|n+1}\kln{x}
%\\
%\label{compx_twist_m_op2_b}
&=&
 \frac{n!}{\kln{n+2-2j}!} \; \int \text{d}^4\!u \; M_{[\rho\beta]}\kln{u} \, \kln{u^2}^{j-1} \; u^{[\rho} x^{\beta]} \;
h^2_{n+1-2j}\kln{u|x} \; , \qquad\qquad\qquad\qquad j\ge 1\,,
\\
\hspace{-.5cm} \label{compx_twist_m_op2_vec}
 M^{\tw\kln{1+2j}\;\text{B}}_{\mu|n+1-2j}\kln{x}
%&=&
% \frac{1}{n+2-2j} \; \pd_\mu \; H_{n+2-2j} \; \Box^{j-1} \; \pd^\rho \; M_{\rho|n+1}\kln{x}
%\\
%\label{compx_twist_m_op2_vec_b}
&=&
 \frac{n!}{\kln{n+2-2j}\kln{n+2-2j}!} \; \int \text{d}^4\!u \; M_{[\rho\beta]}\kln{u} \, \kln{u^2}^{j-1} \; \pd_\mu \; u^{[\rho} x^{\beta]} \;
h^2_{n+1-2j}\kln{u|x} \; ,\quad j \ge 1\,,
\\
\hspace{-.5cm} \label{compx_twist_m_op3}
 M^{\tw\kln{2+2j}\;\text{C}}_{\mu|n+1-2j}\kln{x}
%&=&
%\frac{2}{n+2-2j} \; x^\sigma \kln{ \delta_{[\mu}^\rho \, \pd_{\sigma]} + \frac{1}{n+2-2j} \; \X_{[\mu\sigma]} \; \pd^{\rho} } \; H_{n+1-2j} \;
%\Box^{j} \; M_{\rho|n+1}\kln{x}
%\\
%\label{compx_twist_m_op3_b}
&=&
 \frac{n!}{\kln{n+2-2j}\kln{n+2-2j}!} \; x^\sigma \kln{ \delta_{[\mu}^\rho \pd_{\sigma]} \kln{n+2-2j} + \X_{[\mu\sigma]} \, \pd^\rho } \int
\text{d}^4\!u \; M_{[\rho\beta]}\kln{u} \, \kln{u^2}^{j-1}
\\
\nonumber
&&
 \qquad \times \KLEE{ 4j \, u^\beta \; h^1_{n+1-2j}\kln{u|x} + u^2 \KLn{ 2\,x^\beta \, h^2_{n-2j}\kln{u|x} - u^\beta \, x^2 \,
h^2_{n-1-2j}\kln{u|x} } } .
\end{eqnarray}
%!p____________________________________________________________________________________________________________________________________
In the on-cone limit, $x^2\ra 0$, using $\tx^\sigma \; \tx_{[\mu}
M_{\sigma]|n-1}^{\tw 3\;\text{B}}\kln{\tx} = n/2 \;\, \tx_\mu
M_n^{\tw 3\;\text{B}}\kln{\tx}$, we find two contributions, one of
twist-2 and another one of twist-3,
%!p____________________________________________________________________________________________________________________________________
\begin{equation}
 M_{\mu|n+1}\kln{\tx} = M_{\mu|n+1}^{\tw2\;\text{C}}\kln{\tx} + \frac{\tx_\mu}{n+2} \; M_{n}^{\tw3 \; \text{B}
}\kln{\tx}
\end{equation}
%!p____________________________________________________________________________________________________________________________________
with
%!p____________________________________________________________________________________________________________________________________
\begin{eqnarray}
 M_{\mu|n+1}^{\tw2\;\text{C}}\kln{\tx}
&=&
 \frac{2}{\kln{n+1}\kln{n+2}} \; \tx^\sigma \kln{ \delta_{[\mu}^\rho \, \D_{\sigma]} + \frac{1}{n+2} \; \X_{[\mu\sigma]} \; \D^{\rho} }
M_{\rho|n+1}\kln{\tx}
\\
\nonumber
&=&
 \kln{ \delta_\mu^\rho - \frac{1}{n+2} \; \tx_\mu \, \pd^\rho } \; M_{\rho|n+1}\kln{\tx}
\\
 M_{n}^{\tw3 \; \text{B} }\kln{\tx}
&=&
 \pd^\rho \, M_{\rho|n+1}\kln{\tx} \; .
\end{eqnarray}
%!p____________________________________________________________________________________________________________________________________
Here we have used the relation $\tx^\rho \,
M_{\rho|n+1}\kln{\tx}=0$. After the summation we again compare
with Ref.~\cite{GLR99}. The twist-2 part obtains the form (3.74)
and the twist-3 part is equal to (3.76). In Ref.~\cite{GL01} the
twist-2 part is given by (2.28) and the twist-3 part by (2.29).
Again, we confirm the correct on-cone limit.

To perform the summation to a non-local infinite twist
decomposition we first define non-local operators of well-defined
twist by
%!p____________________________________________________________________________________________________________________________________
\begin{eqnarray}
\label{compx_summe_nlb_m}
 M^{\tw\kln{1+2j}\;\text{B}}\kln{\kappa\,x,-\kappa\,x}
&:=&
 \sum_{n=0}^\infty \; \frac{\im^n}{n!} \; \kappa^{n+2-2j} \; M^{\tw\kln{1+2j}\;\text{B}}_{n+2-2j}\kln{x} \; ,
\\
\label{compx_summe_nlb_mmu_B}
 M_\mu^{\tw\kln{1+2j}\;\text{B}}\kln{\kappa\,x,-\kappa\,x}
&:=&
 \sum_{n=0}^\infty \; \frac{\im^n}{n!} \; \kappa^{n+1-2j} \; M^{\tw\kln{1+2j}\;\text{B}}_{\mu|n+1-2j}\kln{x} \; ,
\\
\label{compx_summe_nlb_mmu_C}
 M_\mu^{\tw\kln{2+2j}\;\text{C}}\kln{\kappa\,x,-\kappa\,x}
&:=&
 \sum_{n=0}^\infty \; \frac{\im^n}{n!} \; \kappa^{n+1-2j} \; M^{\tw\kln{2+2j}\;\text{C}}_{\mu|n+1-2j}\kln{x} \; .
\end{eqnarray}
%!p____________________________________________________________________________________________________________________________________
Again, we have chosen them homogeneous in $\kappa$ and $x$ because
they are related to the (infinite) twist decomposition of the skew
tensor operator $M_{[\mu\nu]}\kln{\kappa\,x,-\kappa\,x}$. After a
straightforward calculation, using the summation properties of
$h^\nu_n\kln{u|x}$, Eq.~(\ref{compx_summe}), and the integral
representation for $1/\kln{n+2-2j}$, we find
%!p____________________________________________________________________________________________________________________________________
\begin{align}
\label{compx_summe_nlb_msc}
 \hspace{-.25cm}
 M^{\tw\kln{1+2j}\;\text{B}}\kln{\kappa\,x,-\kappa\,x}
&=
 \int \text{d}^4\!u \; M_{[\rho\beta]}\kln{u}\; \kln{-u^2}^{j-1} \; \im u^{[\rho} \, \kappa x^{\beta]} \; \kln{2+x\pd} \kln{3+x\pd}
\; \HB^2\kln{u|\kappa\,x} \; , \quad\qquad\quad~ j\ge 1\,,
\\
\hspace{-.25cm}
 \label{compx_summe_nlb_mm}
 M_\mu^{\tw\kln{1+2j}\;\text{B}}\kln{\kappa\,x,-\kappa\,x}
&=
 \int \text{d}^4\!u \; M_{[\rho\beta]}\kln{u}\, \kln{-u^2}^{j-1} \pd_\mu
 \; \im u^{[\rho} \, x^{\beta]} \, \kln{2+x\pd} \kln{3+x\pd}
 \int_0^1 \text{d}\lambda \, \HB^2\kln{u|\lambda\kappa\,x} \,
,\quad j \ge 1\,,
\\
\hspace{-.25cm}
 \label{compx_summe_nlc_mm}
 M^{\tw\kln{2+2j}\;\text{C}}_{\mu}\kln{\kappa\,x,-\kappa\,x}
&=
 \int \text{d}^4\!u \; M_{[\rho\beta]}\kln{u}\; \kln{-u^2}^{j-1} x^\sigma \kln{ \delta_{[\mu}^\rho \pd_{\sigma]}
\kln{1+x\pd} + \X_{[\mu\sigma]} \, \pd^\rho }
\\
 \hspace{-.25cm} \nonumber &
 \times \! \int_0^1 \! \text{d}\lambda \KLEE{ 4j \; \im u^\beta \, \HB^1\kln{u|\lambda\kappa\,x} - u^2 \! \kln{ 2\,\lambda\kappa \, x^\beta
\kln{3+x\pd} - \im u^\beta \! \kln{\lambda\kappa}^2 x^2 } \HB^2\kln{u|\lambda\kappa\,x} } .
\end{align}
%!p____________________________________________________________________________________________________________________________________
Now we get the {\em off-cone representation for the complete
twist-decomposition of} $M_\mu\kln{\kappa\,x,-\kappa\,x}$ as
follows:
%!p____________________________________________________________________________________________________________________________________
%\begin{center}
%\fbox{
%\begin{minipage}[c]{16.5cm}
\begin{eqnarray}
\label{fertig_mmunu} \hspace{-1cm}
 &&
\kappa\, M_{\mu}\kln{\kappa \, x, - \kappa \, x}
 %\\  \nonumber&&
 = M^{\tw2\;\text{C}}_{\mu}\kln{\kappa\,x,-\kappa\,x}
\\
\nonumber \hspace{-1cm} && \qquad\qquad\qquad
 + \; \sum_{j=1}^{\infty} \frac{\kln{\kappa^2 x^2}^{j-1}}{ 4^j \, j! \, \kln{j-1}! } \; \int_0^1
 \frac{\text{d}t}{t} \;
\kln{1-t}^{j-1} \KLEEo{ \kln{t \kappa}^2 x^2 \, M^{\tw\kln{2+2j}\;\text{C}}_{\mu}\kln{t\kappa\,x,-t\kappa\,x} }
\\
\nonumber\hspace{-1cm}
 && \qquad\qquad\qquad
 - \; 4j \, \kln{x\pd} \int_0^1 \text{d}\lambda \, \KLN{ \kln{\lambda t \kappa}^2 x^2 \,
M_{\mu}^{\tw\kln{1+2j} \; \text{B} }\kln{\lambda t\kappa\,x,-\lambda t\kappa\,x}
- \, \kln{\lambda t \kappa} \; x_\mu \; M^{\tw\kln{1+2j} \; \text{B} }\kln{\lambda t\kappa\,x,-\lambda t\kappa\,x} } \oKLEE{} .
\end{eqnarray}
%\end{minipage} }
%\end{center}
%!p____________________________________________________________________________________________________________________________________
This result will be much simpler than the decomposition
(\ref{compx_fertig_nl1}) because we only need the integral
representations (\ref{compx_euler_beta}) and
(\ref{compx_summe_fraction2}) which are both valid starting from
$j=1$. %Let us remark that the homogeneous derivation $\kln{x\pd}$
%could be replaced under the integral by $\lambda \partial_\lambda$
%followed by a partial integration.
%Again, a single integral
%representation for the double integral $\int_0^1\text{d}t \int_0^1
%\text{d} \lambda$ in the non-local expansion leads to general
%hypergeometric functions given by (\ref{hyper_2}) for $n\ra n+1$.

%%%%%%%%%%%%%%%%%%%%%%%%%%%%%%%%%%%%%%%%%%%%%%%%%%%%%%%%%%%%%%%%%%%%%%%%%%%%%%%%%%%%%%%%%%%%%%%%%%%%%%%%%%%%%%%%%%%%%%%%%%%%%%%%%%%%%%
\section{Power corrections of (non)forward matrix elements in $x$--space}
%%%%%%%%%%%%%%%%%%%%%%%%%%%%%%%%%%%%%%%%%%%%%%%%%%%%%%%%%%%%%%%%%%%%%%%%%%%%%%%%%%%%%%%%%%%%%%%%%%%%%%%%%%%%%%%%%%%%%%%%%%%%%%%%%%%%%%

In this sections we are going to consider the implication of our
results to deep inelastic scattering and deeply virtual Compton
scattering. To this purpose we have to take the (non)diagonal
matrix elements of the infinite off-cone decompositions
(\ref{compx_fertig_nl1}) and (\ref{fertig_mmunu}) for the
non-local operators $O_\mu\kln{\kappa\,x,-\kappa\,x}$ and
$M_\mu\kln{\kappa\,x,-\kappa\,x}$, respectively, with the incoming
and outgoing hadron states $\left. |P_1,S_1\right>$ and
$\left<P_2,S_2|\right.$.
%
%In case of the operator $O_\mu\kln{\zeta\,x,-\zeta\,x}$ these are the two scalar operators
%$O^{\tw\kln{2+2j}\;\text{A/B}}\kln{\zeta\,x,-\zeta\,x}$, the two related vector operators
%$O_\mu^{\tw\kln{2+2j}\;\text{A/B}}\kln{\zeta\,x,-\zeta\,x}$ and the operator
%$O_\mu^{\tw\kln{3+2j}\;\text{C}}\kln{\zeta\,x,-\zeta\,x}$.
%For $M_\mu\kln{\zeta\,x,-\zeta\,x}$ we have to deal with $M^{\tw\kln{1+2j}\;\text{B}}\kln{\zeta\,x,-\zeta\,x}$,
%$M_\mu^{\tw\kln{1+2j}\;\text{B}}\kln{\zeta\,x,-\zeta\,x}$ and
%$M_\mu^{\tw\kln{2+2j}\;\text{C}}\kln{\zeta\,x,-\zeta\,x}$.
%
Before doing this, again separately for the various operators of
well-defined twist $\matel{P_2,S_2}{\OP_\Gamma^{\tw\kln{\tau}}
\kln{\zeta\,x,-\zeta\,x}}{P_1,S_1}$, let us shortly review the
parametrization of such matrix elements as it has been introduced
in Ref.~\cite{GLR01}. Here, $\Gamma$ denotes the Dirac structure
of the operator and $\zeta$ denotes any of the different products
of $\lambda$, $t$ and $\kappa$ appearing in the arguments of the
non-local operators.

%For deep inelastic scattering we have $P_1=P_2$ and $S_1=S_2$.
%$\OP$ again generically denotes $O$ and $M$ and $\Gamma$ the two
%different $\Gamma$-structures $\gamma_\mu$ and $x^\nu \,
%\sigma_{\mu\nu}$. The twist $\tau$ is regarded as fixed but not
%further specified which means that the operators
%$\OP_\Gamma^{\tw\kln{\tau}}\kln{\zeta\,x,-\zeta\,x}$ can be of
%type A, B or C.

According to that approach the matrix elements of operators of
well-defined twist are parametrized by {\em Lorentz invariant
double distributions} $f_a^{\kln{\tau}}\kln{\Z,\mu^2}$, not
suffering from any power corrections, and suitable {\em
kinematical factors} $\KI^a_\Gamma\kln{P_1,P_2;S_1,S_2}$ referring
to the $\Gamma$-structure of the considered operator on the
hadronic level:
%!p____________________________________________________________________________________________________________________________________
\begin{eqnarray}
\label{para1}
\matel{P_2,S_2}{\OP_\Gamma^{\tw\kln{\tau}}\kln{\zeta\,x,-\zeta\,x}}{P_1,S_1}
&\equiv& \matel{P_2,S_2}{\PRa^{\kln{\tau}\Gamma'}_{\Gamma}
\OP_{\Gamma'}^{\tw\kln{\tau}}\kln{\zeta\,x,-\zeta\,x}}{P_1,S_1}
\nonumber\\
 &=& \PRa^{\kln{\tau}\Gamma'}_{\Gamma}(x,\partial) \,
\KI^a_{\Gamma'}\kln{\PP, \mathbb{S}} \, \int \text{D}\Z \;
\e^{\im\kappa\kln{x\PP}\Z} \; f_a^{\kln{\tau}}\kln{\Z,\mu^2} \; .
\end{eqnarray}
%!p____________________________________________________________________________________________________________________________________
Here, $\PRa^{\kln{\tau}\Gamma'}_{\Gamma}(x,\partial)$ are the
corresponding projections onto operators of well-defined twist
$\tau$,
\begin{eqnarray}
\OP_{\Gamma}^{\tw\kln{\tau}}\kln{\zeta\,x,-\zeta\,x} =
\PRa^{\kln{\tau}\Gamma'}_{\Gamma}(x,\partial)
\;\OP_{\Gamma'}\kln{\zeta\,x,-\zeta\,x}\,,
\end{eqnarray}
which follow from their local versions in
Eqs.~(\ref{compx_twist_op1}) -- (\ref{compx_twist_op3_vec}) in the
form (\ref{compx_summea}) -- (\ref{compx_summec}) and
(\ref{compx_summe_nlb_msc}) -- (\ref{compx_summe_nlc_mm}),
respectively, eventually including integrations over $\lambda$.
In the representation (\ref{para1}) these projection operators act
only on the $\Gamma$--structure and the exponential
$\e^{\im\kappa\kln{x\PP}\Z}$. In addition, we have used the
notation $\PP=\kls{P_+,P_-}$ and $\Z=\kls{z_+,z_-}$ with
$P_\pm=P_2\pm P_1$ and $z_\pm=1/2\kln{z_2\pm z_1}$. The
integration measure is given by
$\text{D}\Z=\text{d}z_1\,\text{d}z_2 \;
\theta\kln{1-z_1}\theta\kln{z_1+1}\theta\kln{1-z_2}\theta\kln{z_2+1}$
and reflects the fact that the double distributions
$f_a^{\kln{\tau}}\kln{x\PP,\mu^2}$ are entire analytic functions
in $x\PP$-space. We assumed summation convention with respect to
$a$. For deep inelastic scattering we have only one momentum $P$
and one distribution variable $z$.

Applying this parametrization, after the formal Fourier
transformations (\ref{compx_momo}) and (\ref{compx_momm}), the
replacements
%!p____________________________________________________________________________________________________________________________________
\begin{eqnarray}
\label{replacement_deep}
 \Matel{P,S}{\OP_\Gamma\kln{u}}{P,S}
&\overset{\cdot}{=}& \KI^a_{\Gamma}\kln{P,S} \int \text{d}z \;\,
\delta^{\kln{4}}\kln{u-2\,Pz} \, \hat
f_a^{\kln{\tau}}\kln{z,\mu^2} \; ,
\\
\label{replacement_virtual}
\Matel{P_2,S_2}{\OP_\Gamma\kln{u}}{P_1,S_1} &\overset{\cdot}{=}&
\KI^a_{\Gamma}\kln{\PP,\mathbb{S}} \int \text{D}\Z \;\,
\delta^{\kln{4}}\kln{u-\PZ} \, f_a^{\kln{\tau}}\kln{\Z,\mu^2} \;,
\end{eqnarray}
%!p____________________________________________________________________________________________________________________________________
are deduced, which hold under the application of the twist
projection $\PRa^{\kln{\tau}\Gamma'}_{\Gamma}$. For a detailed
discussion of (\ref{replacement_deep}) and
(\ref{replacement_virtual}), see, Ref.~\cite{GLR01}.

%%%%%%%%%%%%%%%%%%%%%%%%%%%%%%%%%%%%%%%%%%%%%%%%%%%%%%%%%%%%%%%%%%%%%%%%%%%%%%%%%%%%%%%%%%%%%%%%%%%%%%%%%%%%%%%%%%%%%%%%%%%%%%%%%%%%%%
\subsection{Power corrections of forward matrix elements}
%%%%%%%%%%%%%%%%%%%%%%%%%%%%%%%%%%%%%%%%%%%%%%%%%%%%%%%%%%%%%%%%%%%%%%%%%%%%%%%%%%%%%%%%%%%%%%%%%%%%%%%%%%%%%%%%%%%%%%%%%%%%%%%%%%%%%%

In the case of deep inelastic scattering we have $P_1=P_2$ and
$S_1=S_2$. For the operator $O_\mu\kln{\kappa\,x,-\kappa\,x}$
there remains only the Dirac structure $\KI^1_\mu = 2P_\mu$, see
also Ref.~\cite{GL01}. Then, after inserting the replacement
(\ref{replacement_deep}), for the two non-local scalar operators
(\ref{compx_summea}) and (\ref{compx_summeb}) we obtain the
following forward matrix elements:
%!p____________________________________________________________________________________________________________________________________
\begin{align}
\label{compx_summea_matel}
&
 \matel{P}{O^{\tw\kln{2+2j}\;\text{A}}\kln{\zeta\,x,-\zeta\,x}}{P}
 \\
 \nonumber
& \qquad\qquad
 = \, 2 \, P_\rho \int \text{d}z \; \hat F_A^{\kln{2+2j}}\kln{z} \; \kln{-\kln{2\,Pz}^2}^{j-1}
 \KLEEo{} 4j \;\, \im z \, P^{\rho}\, \kln{ 1 + x\pd } \; \HB^1\kln{2\,Pz|\zeta \, x}
\\
\nonumber
&
 \qquad \qquad \qquad \qquad \qquad\qquad
  - \, \kln{2\,Pz}^2 \KLn{ \zeta \, x^\rho \, \kln{2+x\pd}
  - \im z \, P^{\rho}\; \zeta^2 \, x^2 } \kln{3+x\pd} \;
\HB^2\kln{2\,Pz|\zeta \, x} \oKLEE{} \; ,
\\
\label{compx_summeb_matel}
&
 \matel{P}{O^{\tw\kln{2+2j}\;\text{B}}\kln{\zeta\,x,-\zeta\,x}}{P}
 \\
 \nonumber
& \qquad\qquad
 = 4\, P_\rho \int \text{d}z \; \hat
F_B^{\kln{2+2j}}\kln{z} \;
 \kln{-\kln{2\,Pz}^2}^{j-1} \, \im z \, P^{\rho}\, \kln{1+x\pd} \; \HB^1\kln{2\,Pz|\zeta \, x} \; .
\\ %\end{eqnarray}
%!p____________________________________________________________________________________________________________________________________
\intertext{The matrix elements of the two related vector operators
(\ref{compx_summea_vec}) and (\ref{compx_summeb_vec}) obtain a
quite similar form:}
%!p____________________________________________________________________________________________________________________________________
%\begin{eqnarray}
\label{compx_summea_vec_matel}
&
 \matel{P}{O_\mu^{\tw\kln{2+2j}\;\text{A}}\kln{\zeta\,x,-\zeta\,x}}{P}
 \\
\nonumber & \qquad\qquad
 = \frac{2}{\zeta} \; P_\rho \int
\text{d}z \; \hat F_A^{\kln{2+2j}}\kln{z} \;
\kln{-\kln{2\,Pz}^2}^{j-1}
 \pd_\mu \; \int_0^1 \frac{\text{d}\lambda}{\lambda}
\KLEEo{} 4 j \;\, \im z \, P^{\rho}\, \kln{ 1 + x\pd } \;
\HB^1\kln{2\,Pz| \lambda \zeta \, x}
\\
\nonumber
&
 \qquad \qquad \qquad \qquad \qquad \qquad\qquad
 - \, \kln{2\,Pz}^2 \KLn{ \lambda\zeta \; x^\rho \, \kln{2+x\pd}
 - \im z \, P^{\rho}\; \kln{\lambda \zeta}^2 \, x^2 } \kln{3+x\pd} \;
\HB^2\kln{2\,Pz| \lambda \zeta \, x} \oKLEE{} \; ,
\\
\label{compx_summeb_vec_matel}
&
 \matel{P}{O_\mu^{\tw\kln{2+2j}\;\text{B}}\kln{\zeta\,x,-\zeta\,x}}{P}
 \\
 \nonumber
& \qquad\qquad
 = \frac{4}{\zeta} \; P_\rho \int \text{d}z \; \hat F_B^{\kln{2+2j}}\kln{z} \;
 \kln{-\kln{2\,Pz}^2}^{j-1}
 \im z \, P^{\rho}\;\; \pd_\mu \; \kln{1+x\pd} \int_0^1 \frac{\text{d}\lambda}{\lambda} \;
 \HB^1\kln{2\,Pz| \lambda \zeta \, x} \; .
\\ %\end{eqnarray}
%!p____________________________________________________________________________________________________________________________________
\intertext{The last matrix element, giving all contributions of
odd twist being contained in the operator
$O_\mu\kln{\kappa\,x,-\kappa\,x}$, reads}
%!p____________________________________________________________________________________________________________________________________
%\begin{eqnarray}
\label{compx_summec_matel}
&
 \matel{P}{O_\mu^{\tw\kln{3+2j}\;\text{C}}\kln{\zeta\,x,-\zeta\,x}}{P}
 \\
 \nonumber
& \qquad\qquad
 = 4 \, P_\rho \int \text{d}z \; \hat
F_C^{\kln{3+2j}}\kln{z} \; \kln{-\kln{2\,Pz}^2}^{j} x^\sigma \kln{
\delta_{[\mu}^\rho \pd_{\sigma]} \kln{ 1 + x\pd } +
\X_{[\mu\sigma]} \, \pd^\rho } \int_0^1 \text{d} \lambda \;\,
\HB^1\kln{2\,Pz|\lambda \zeta \, x} \; .
\end{align}
%!p____________________________________________________________________________________________________________________________________
All these matrix elements are written down for unpolarized
scattering, where only the momentum $P_\mu$ can be used for the
construction of kinematical factors.
%It is therefore not possible to build the antisymmetric matrix element
%$\matel{P}{M_{[\rho\beta]}\kln{u}}{P}$ which is needed for the
%construction of unpolarized forward matrix elements of the
%operators (\ref{compx_summe_nlb_msc}) to (\ref{compx_summe_nlc_mm}).

The operator $M^5_\mu\kln{\kappa\,x,-\kappa\,x}$ only contributes
to polarized deep inelastic scattering. Here, we have to use the
kinematical factor $\KI^1_{[\mu\nu]}\kln{P,S} = 2 \kln{S_\mu P_\nu
- S_\nu P_\mu}/M$. The three related matrix elements read
%!p____________________________________________________________________________________________________________________________________
\begin{align}
\label{compx_summe_nlb_msc_matel}
 &
 \matel{PS}{M^{5\;\tw\kln{1+2j}\;\text{B}}\kln{\zeta\,x,-\zeta\,x}}{PS}
 \\
 \nonumber
& \qquad
 = \frac{4}{M} \; \kln{S_\rho P_\beta - S_\beta P_\rho}
 \int \text{d}z \; \hat H_B^{\kln{1+2j}}\kln{z} \; \kln{-\kln{2\,Pz}^2}^{j-1} \; \im z \, P^{[\rho} \, \zeta x^{\beta]} \; \kln{2+x\pd} \kln{3+x\pd}
\; \HB^2\kln{2\,Pz|\zeta\,x}
 \\
 \nonumber
& \qquad
 = - \, 4\im \, \zeta \, M \kln{xS}
 \int \text{d}z \; z \, \hat H_B^{\kln{1+2j}}\kln{z} \; \kln{-4 \, z^2 M^2}^{j-1} \kln{2+x\pd} \kln{3+x\pd}
\; \HB^2\kln{2\,Pz|\zeta\,x} \; ,
\\
%\intertext{}
 \label{compx_summe_nlb_mm_matel}
 &
 \matel{PS}{M_\mu^{5\;\tw\kln{1+2j}\;\text{B}}\kln{\zeta\,x,-\zeta\,x}}{PS}
 \\
 \nonumber
& \qquad
 = \frac{4}{M} \; \kln{S_\rho P_\beta - S_\beta P_\rho}
 \int \text{d} z \; \hat H_B^{\kln{1+2j}}\kln{z} \; \kln{-\kln{2\,Pz}^2}^{j-1} \pd_\mu \; \im z \, P^{[\rho} \, x^{\beta]} \; \kln{2+x\pd} \kln{3+x\pd}
\; \int_0^1 \text{d}\lambda \; \HB^2\kln{2\,Pz|\lambda\zeta\,x}
 \\
 \nonumber
& \qquad
 = - \, 4\im \, M
 \int \text{d} z \; z \, \hat H_B^{\kln{1+2j}}\kln{z} \; \kln{-4 \, z^2 M^2}^{j-1} \pd_\mu \, \kln{xS} \, \kln{2+x\pd} \kln{3+x\pd}
\; \int_0^1 \text{d}\lambda \; \HB^2\kln{2\,Pz|\lambda\zeta\,x} \;
,
\\
%\intertext{}
 \label{compx_summe_nlc_mm_matel} &
 \matel{PS}{M_\mu^{5\;\tw\kln{2+2j}\;\text{C}}\kln{\zeta\,x,-\zeta\,x}}{PS}
 \\
 \nonumber
& \qquad
 = \frac{4}{M} \; \kln{S_\rho P_\beta - S_\beta P_\rho}
 \int \text{d}z \; \hat H_C^{\kln{2+2j}}\kln{z} \; \kln{-\kln{2\,Pz}^2}^{j-1} x^\sigma \kln{ \delta_{[\mu}^\rho \pd_{\sigma]}
\kln{1+x\pd} + \X_{[\mu\sigma]} \, \pd^\rho }
\\
\nonumber & \qquad \qquad
 \times \! \int_0^1 \! \text{d}\lambda \KLEE{ 4j \; \im z \, P^{\beta} \, \HB^1\kln{2\,Pz|\lambda\zeta\,x} - \kln{2\,Pz}^2 \! \kln{ \lambda\zeta \, x^\beta
\kln{3+x\pd} - \im z \, P^{\beta} \! \kln{\lambda\zeta}^2 x^2 }
\HB^2\kln{2\,Pz|\lambda\zeta\,x} } .
\end{align}
%!p____________________________________________________________________________________________________________________________________
where we have used $S\,P=0$ and $P^2=M^2$.

If we want to take polarized matrix elements of the operator
$O^5_\mu\kln{\kappa\,x,-\kappa\,x}$ we have to use the replacement
%!p____________________________________________________________________________________________________________________________________
\begin{eqnarray}
\label{replacement_deep_pol}
\Matel{PS}{O^5_\rho\kln{u}}{PS}
&\overset{\cdot}{=}&
2 \, S_\rho \int \text{d}z \;\, \delta^{\kln{4}}\kln{u-2\,Pz} \, \hat G_a^{\kln{\tau}}\kln{z,\mu^2}
\end{eqnarray}
%!p____________________________________________________________________________________________________________________________________
under the different twist projectors. The resulting matrix
elements will be of the form (\ref{compx_summea_matel}) to
(\ref{compx_summec_matel}) with $P_\rho$ replaced by $S_\rho$ and
$\hat F$ replaced by $\hat G$.
%We therefore do not repeat these results.

Forward matrix elements on the light-cone of non-local operators
of well-defined twist have been constructed in Ref.~\cite{GL01}
for a large number of operators. If we perform the on-cone limit
of our results (\ref{compx_summea_matel}) --
(\ref{compx_summe_nlc_mm_matel}) we again find the correct on-cone
limits.
%We give no detailed calculations for this limit but as an
%example we instead refer to ref.\cite{GLR01} page 110 where the
%on-cone limit for the matrix element (\ref{compx_summea_matel})
%for $j=0$ is performed.

%%%%%%%%%%%%%%%%%%%%%%%%%%%%%%%%%%%%%%%%%%%%%%%%%%%%%%%%%%%%%%%%%%%%%%%%%%%%%%%%%%%%%%%%%%%%%%%%%%%%%%%%%%%%%%%%%%%%%%%%%%%%%%%%%%%%%%
\subsection{Power corrections of non-forward matrix elements}
%%%%%%%%%%%%%%%%%%%%%%%%%%%%%%%%%%%%%%%%%%%%%%%%%%%%%%%%%%%%%%%%%%%%%%%%%%%%%%%%%%%%%%%%%%%%%%%%%%%%%%%%%%%%%%%%%%%%%%%%%%%%%%%%%%%%%%

To apply the parametrization (\ref{replacement_virtual}) to the
operators $O_\mu\kln{\kappa\,x,-\kappa\,x}$ and
$O^5_\mu\kln{\kappa\,x,-\kappa\,x}$ we have to fix the kinematical
factors $\KI^{(5)\,a}_{\Gamma}\kln{\PP,\mathbb{S}}$ which are
given by the Dirac and Pauli structures,
%!p____________________________________________________________________________________________________________________________________
\begin{eqnarray}
\label{dirac} \KI^{(5)\;1}_{\mu}\kln{\PP,\mathbb{S}} &=& \bar
u\kln{P_2,S_2} \kln{\gamma^5} \gamma_\mu \, u\kln{P_1,S_1} \; ,
\\
\label{pauli} \KI^{(5)\;2}_{\mu}\kln{\PP,\mathbb{S}} &=&
\frac{1}{M} \; \bar u\kln{P_2,S_2} \kln{\gamma^5} \sigma_{\mu\nu}
\, P_-^\nu \;\, u\kln{P_1,S_1} \; ,
\end{eqnarray}
%!p____________________________________________________________________________________________________________________________________
(see also Ref.~\cite{BGR99}) and then insert the replacement
(\ref{replacement_virtual}) into the different matrix elements.

The non-forward matrix elements of the two non-local scalar
operators (\ref{compx_summea}) and (\ref{compx_summeb}) obtain the
form
%!p____________________________________________________________________________________________________________________________________
\begin{align}
\label{compx_summea_matel_virtual}
&
 \matel{P_2,S_2}{O^{\tw\kln{2+2j}\;\text{A}}\kln{\zeta\,x,-\zeta\,x}}{P_1,S_1}
 \\
 \nonumber
& \qquad\qquad
 = \frac{1}{2} \, \KI^a_{\rho}\kln{\PP,\mathbb{S}} \int \text{D}\Z \; F_{Aa}^{\kln{2+2j}}\kln{\Z} \; \kln{-\kln{\PZ}^2}^{j-1}
 \KLEEo{} 4j \;\, \im \PZi{\rho} \, \kln{ 1 + x\pd } \; \HB^1\kln{\PZ|\zeta \, x}
\\
\nonumber
&
 \qquad \qquad \qquad \qquad \qquad\qquad
  - \, \kln{\PZ}^2 \KLn{ 2 \, \zeta \, x^\rho \, \kln{2+x\pd} - \im \PZi{\rho} \;
  \zeta^2 \, x^2 } \kln{3+x\pd} \;
\HB^2\kln{\PZ|\zeta \, x} \oKLEE{} \; ,
\\
\label{compx_summeb_matel_virtual}
&
 \matel{P_2,S_2}{O^{\tw\kln{2+2j}\;\text{B}}\kln{\zeta\,x,-\zeta\,x}}{P_1,S_1}
 \\
 \nonumber
& \qquad\qquad
 = \KI^a_{\rho}\kln{\PP,\mathbb{S}
} \int \text{D}\Z \; F_{Ba}^{\kln{2+2j}}\kln{\Z} \;
 \kln{-\kln{\PZ}^2}^{j-1} \, \im \PZi{\rho} \, \kln{1+x\pd} \; \HB^1\kln{\PZ|\zeta \, x} \; .
\\ %\end{eqnarray}
%!p____________________________________________________________________________________________________________________________________
\intertext{The matrix elements of the two related vector operators
(\ref{compx_summea_vec}) and (\ref{compx_summeb_vec}) again obtain
the similar form}
%!p____________________________________________________________________________________________________________________________________
%\begin{eqnarray}
\label{compx_summea_vec_matel_virtual}
&
 \matel{P_2,S_2}{O_\mu^{\tw\kln{2+2j}\;\text{A}}\kln{\zeta\,x,-\zeta\,x}}{P_1,S_1}
 \\
\nonumber
& \qquad\qquad
 = \frac{1}{2\,\zeta} \;
\KI^a_{\rho}\kln{\PP,\mathbb{S}} \int \text{D}\Z \;
F_{Aa}^{\kln{2+2j}}\kln{\Z} \; \kln{-\kln{\PZ}^2}^{j-1}
 \pd_\mu \; \int_0^1 \frac{\text{d}\lambda}{\lambda}
\KLEEo{} 4 j \;\, \im \PZi{\rho} \, \kln{ 1 + x\pd } \;
\HB^1\kln{\PZ| \lambda \zeta \, x}
\\
\nonumber
&
 \qquad \qquad \qquad \qquad \qquad \qquad\qquad
 - \, \kln{\PZ}^2 \KLn{ 2 \, \lambda\zeta \; x^\rho \, \kln{2+x\pd} - \im \PZi{\rho} \; \kln{\lambda \zeta}^2 \, x^2 } \kln{3+x\pd} \;
\HB^2\kln{\PZ| \lambda \zeta \, x} \oKLEE{} \; ,
\\
\label{compx_summeb_vec_matel_virtual}
&
 \matel{P_2,S_2}{O_\mu^{\tw\kln{2+2j}\;\text{B}}\kln{\zeta\,x,-\zeta\,x}}{P_1,S_1}
 \\
 \nonumber
& \qquad\qquad
 = \frac{1}{\zeta} \; \KI^a_{\rho}\kln{\PP,\mathbb{S}} \int \text{D}\Z \; F_{Ba}^{\kln{2+2j}}\kln{\Z} \; \kln{-\kln{\PZ}^2}^{j-1}
 \im \PZi{\rho} \;\; \pd_\mu \; \kln{1+x\pd} \int_0^1 \frac{\text{d}\lambda}{\lambda} \;
 \HB^1\kln{\PZ| \lambda \zeta \, x} \; .
\\ %\end{eqnarray}
%!p____________________________________________________________________________________________________________________________________
\intertext{The last matrix element of the type C vector operator
(\ref{compx_summec}) gives all contribution of odd twist for
unpolarized deeply virtual Compton scattering}
%!p____________________________________________________________________________________________________________________________________
%\begin{eqnarray}
\label{compx_summec_matel_virtual}
&
 \matel{P_2,S_2}{O_\mu^{\tw\kln{3+2j}\;\text{C}}\kln{\zeta\,x,-\zeta\,x}}{P_1,S_1}
 \\
 \nonumber
&
 \qquad\qquad
  = 2 \, \KI^a_{\rho}\kln{\PP,\mathbb{S}} \int \text{D}\Z \;
F_{Ca}^{\kln{3+2j}}\kln{\Z} \; \kln{-\kln{\PZ}^2}^{j} x^\sigma
\kln{ \delta_{[\mu}^\rho \pd_{\sigma]} \kln{ 1 + x\pd } +
\X_{[\mu\sigma]} \, \pd^\rho } \int_0^1 \text{d} \lambda \;\,
\HB^1\kln{\PZ|\lambda \zeta \, x} \; .
\end{align}
%!p____________________________________________________________________________________________________________________________________
A parametrization for polarized scattering is found by the
replacement of $\KI^a_{\rho}\kln{\PP,\mathbb{S}}$ by
$\KI^{5\;a}_{\rho}\kln{\PP,\mathbb{S}}$.

For a parametrization of the operator
$M_\mu\kln{\kappa\,x,-\kappa\,x}$ for polarized and unpolarized
deeply virtual Compton scattering we use the kinematical factors
%!p____________________________________________________________________________________________________________________________________
\begin{eqnarray}
\label{para_M_1} \KI^{(5)\;1}_{[\mu\nu]}\kln{\PP,\mathbb{S}} &=&
\frac{1}{M} \; \bar u\kln{P_2,S_2} \kln{\gamma^5} \gamma_{[\mu}
P^+_{\nu]} \, u\kln{P_1,S_1} \; ,
\\
\label{para_M_2} \KI^{(5)\;2}_{[\mu\nu]}\kln{\PP,\mathbb{S}} &=&
\frac{1}{M} \; \bar u\kln{P_2,S_2} \kln{\gamma^5} \gamma_{[\mu}
P^-_{\nu]} \, u\kln{P_1,S_1} \, ,
\end{eqnarray}
%!p____________________________________________________________________________________________________________________________________
and find the matrix elements
%!p____________________________________________________________________________________________________________________________________
\begin{align}
\label{compx_summe_nlb_msc_virtual_matel}
&
 \matel{P_2,S_2}{M^{(5)\;\tw\kln{1+2j}\;\text{B}}\kln{\zeta\,x,-\zeta\,x}}{P_1,S_1}
 \\
 \nonumber
& \qquad\qquad
 = \KI^{(5)\;a}_{[\rho\beta]}\kln{\PP,\mathbb{S}} \int \text{D}\Z \; H_{Ba}^{\kln{1+2j}}\kln{\Z} \; \kln{-\kln{\PZ}^2}^{j-1} \;
 \im \PZi{[\rho} \;\, \zeta x^{\beta]} \; \kln{2+x\pd} \kln{3+x\pd} \; \HB^2\kln{\PZ|\zeta\,x} \; ,
\\
%\intertext{}
 \label{compx_summe_nlb_mm_virtual_matel} &
 \matel{P_2,S_2}{M_\mu^{(5)\;\tw\kln{1+2j}\;\text{B}}\kln{\zeta\,x,-\zeta\,x}}{P_1,S_1}
 \\
 \nonumber
& \qquad\qquad
 = \KI^{(5)\;a}_{[\rho\beta]}\kln{\PP,\mathbb{S}} \int \text{D}\Z \; H_{Ba}^{\kln{1+2j}}\kln{\Z} \; \kln{-\kln{\PZ}^2}^{j-1} \pd_\mu \;
 \im \PZi{[\rho} \;\, x^{\beta]} \; \kln{2+x\pd} \kln{3+x\pd}
\; \int_0^1 \text{d}\lambda \; \HB^2\kln{\PZ|\lambda\zeta\,x} \; ,
\\
%\intertext{}
 \label{compx_summe_nlc_mm_virtual_matel} &
 \matel{P_2,S_2}{M_\mu^{(5)\;\tw\kln{2+2j}\;\text{C}}\kln{\zeta\,x,-\zeta\,x}}{P_1,S_1}
 \\
 \nonumber
& \qquad\qquad
 = \KI^{(5)\;a}_{[\rho\beta]}\kln{\PP,\mathbb{S}} \int \text{D}\Z \; H_{Ca}^{\kln{2+2j}}\kln{\Z} \; \kln{-\kln{\PZ}^2}^{j-1} x^\sigma \kln{ \delta_{[\mu}^\rho \pd_{\sigma]}
\kln{1+x\pd} + \X_{[\mu\sigma]} \, \pd^\rho }
\\
\nonumber & \qquad \qquad\qquad
 \times \! \int_0^1 \! \text{d}\lambda \KLEE{ 4j \; \im \PZi{\beta} \;\; \HB^1\kln{\PZ|\lambda\zeta\,x} - \kln{\PZ}^2 \kln{ 2\,\lambda\zeta \, x^\beta
\kln{3+x\pd} - \im \PZi{\beta} \, \kln{\lambda\zeta}^2 x^2 } \HB^2\kln{\PZ|\lambda\zeta\,x} } .
\end{align}
%!p____________________________________________________________________________________________________________________________________
Notice that the kinematical factor
$\KI^{(5)\;2}_{[\mu\nu]}\kln{\PP,\mathbb{S}}$ vanishes for forward
scattering since $P_-$ vanishes in this case. For unpolarized deep
inelastic scattering $\KI^{1}_{[\mu\nu]}\kln{P}$ also gives zero
since $\bar u\kln{P,S} \gamma_{[\mu} P^+_{\nu]} \, u\kln{P,S} = 4
\, P_{[\mu} P_{\nu]} = 0$. For polarized deep inelastic scattering
we find $\KI^{5\;1}_{[\mu\nu]}\kln{P,S} = 2 \kln{ S_\mu P_\nu -
S_\nu P_\mu }/M$.

%%%%%%%%%%%%%%%%%%%%%%%%%%%%%%%%%%%%%%%%%%%%%%%%%%%%%%%%%%%%%%%%%%%%%%%%%%%%%%%%%%%%%%%%%%%%%%%%%%%%%%%%%%%%%%%%%%%%%%%%%%%%%%%%%%%%%%
\section{Conclusion}
%%%%%%%%%%%%%%%%%%%%%%%%%%%%%%%%%%%%%%%%%%%%%%%%%%%%%%%%%%%%%%%%%%%%%%%%%%%%%%%%%%%%%%%%%%%%%%%%%%%%%%%%%%%%%%%%%%%%%%%%%%%%%%%%%%%%%%
In this paper we introduced a procedure which allows to determine
the decomposition of arbitrary non-local vector operators into an
infinite sum of non-local vector operators of arbitrary twist. The
procedure generalizes the much simpler decomposition of non-local
scalar operators introduced in Ref.~\cite{GLR01}. As examples we
considered the bilocal operators
$O^{(5)}_\mu\kln{\kappa\,x,-\kappa\,x}$ and
$M_\mu\kln{\kappa\,x,-\kappa\,x}= x^\nu
M_{[\mu\nu]}\kln{\kappa\,x,-\kappa\,x}$ together with their
forward and non-forward matrix elements. The latter operator has
been chosen because it shows some peculiarities which are related
to its `internal' antisymmetry and the `external' contraction with
$x^\nu$. However, the procedure is much more general applying to
any kind of non-local vector fields. Other examples could be
purely gluonic bilocal operators or trilocal operators like $x^\nu
\bar \psi (\kappa_1 x) F_{[\mu\nu]}(\kappa_2x) \psi(\kappa_3 x)$
where $F_{[\mu\nu]}$ is the gluon field strength. (In the case of
tri- and multi-local operators one has to apply a prescription of
local operators which was introduced in \cite{GLR00}, Chapter 4.)
Furthermore, let us remark that the trace decomposition as a
prerequisite of that approach - which, in fact, is more
complicated than the application of the symmetry projections - is
also separately of interest.

The procedure introduced here can be extended straightforwardly to
antisymmetric and symmetric tensor operators of rank 2. Thereby, a
hard part is the determination of the construction of traceless
local operators $T_{\mu\nu|n}(x) =
T_{\mu\nu\alpha_1\ldots\alpha_n} x^{\alpha_1} \cdots x^{\alpha_n}$
as well as the implementation of the various symmetry types I to
IV through the differential operators corresponding to
$\Y^{i}_{[m]},\,i= 1,...,4$.

This work is motivated by the aim to study the target mass
contributions to the various QCD-distribution amplitudes which
occur in the phenomenological considerations of hard hadronic
scattering processes, like deep inelastic scattering and deep
virtual Compton scattering, as well as in hadronic form factors.
In the case of hadronic wave functions, e.g. for the pion or the
$\rho-$meson, the obtained twist decompositions, after
corresponding (anti-) symmetrization w.r.t. $\kappa$, could be
used for their power (resp. mass) corrections. However, to give a
comprehensive description the complete twist decomposition of
$M_{[\mu\nu]}\kln{\kappa\,x,-\kappa\,x}$ had to be known. On the
other hand, in order to determine the target mass corrections of
deep virtual Compton scattering in terms of $M^2/Q^2$, as has been
done in Ref.~\cite{BM01} up to twist-3, the Fourier transform of
the (anti-) symmetrized operators of definite twist with the
corresponding coefficient functions according to
Eqs.~(\ref{CA_nonf}) and (\ref{str_wick}) has to be performed. In
the case of the scalar operators this has been done in
\cite{EG02}, and in the case of vector operators this is under
consideration.

\acknowledgments

\noindent The authors are grateful to D. Robaschik for %various
useful discussions. In addition, J. Eilers gratefully acknowledges
the Graduate College "Quantum field theory" at Center for
Theoretical Studies of Leipzig University for financial support.

%\newpage

\end{document}